\documentclass[journal]{IEEEtran}

\ifCLASSINFOpdf
\else
\fi

\usepackage[utf8]{inputenc} 
\usepackage[T1]{fontenc}    
\usepackage{hyperref}       
\usepackage{url}            
\usepackage{booktabs}       
\usepackage{amsfonts}       
\usepackage{nicefrac}       
\usepackage{microtype}      
\usepackage{amsmath,amsthm,amssymb,amsfonts}
\usepackage{amsmath,graphicx}
 \usepackage{xcolor}
\usepackage{algorithm,xpatch}
\usepackage{hyperref}
\usepackage{comment}
\usepackage{float}
   \usepackage{xcolor}
    \usepackage{lipsum}
    \usepackage{amsmath}
\usepackage{algorithm}
\usepackage{multirow}
\usepackage{tabularx} 

\usepackage[noend]{algpseudocode}
\usepackage{amsthm}
\usepackage{enumitem}
\usepackage{pdfpages}
\def\proj{\mathop{\rm proj}\nolimits}
\def\corr{\mathop{\rm corr}\nolimits}
\def\trace{\mathop{\rm trace}\nolimits}
\def\descent{\mathop{\rm descent}\nolimits}
\usepackage{caption} 
\usepackage{xcolor}
 \usepackage{tikz}
\usetikzlibrary{arrows.meta,automata,positioning}
\usepackage[edges]{forest}
\DeclareMathAlphabet{\mathcal}{OMS}{cmsy}{m}{n}
\SetMathAlphabet{\mathcal}{bold}{OMS}{cmsy}{b}{n}
\newcommand{\bigO}{\mathcal{O}}
\newcommand\hl[1]{%
  \bgroup
  \hskip0pt\color{red!80!black}%
  #1%
  \egroup
}

\usepackage{xcolor}

\usepackage[most]{tcolorbox}
\newtcolorbox{highlighted}{colback=yellow,coltext=red,breakable}

\usepackage{caption,subcaption}
 \usepackage{tikz}
\usetikzlibrary{arrows.meta,automata,positioning}

\begin{document}
%
\title{Hierarchical extraction of  functional connectivity components in human brain using resting-state fMRI}
%
%
%

\author{Dushyant Sahoo, Theodore D. Satterthwaite, Christos Davatzikos
\thanks{Dushyant Sahoo and Christos Davatzikos are with Department of Electrical and Systems Engineering University of Pennsylvania, PA, USA (e-mail: sadu@seas.upenn.edu; Christos.Davatzikos@uphs.upenn.edu)}
\thanks{Theodore D. Satterthwaite is with Department of Psychiatry, University of Pennsylvania, PA, USA (email: sattertt@pennmedicine.upenn.edu ) }

}

\maketitle

%

\begin{abstract}
The study of functional networks of the human brain has been of significant interest in cognitive neuroscience for over two decades, albeit they are typically extracted at a single scale using various methods, including decompositions like ICA. However, since numerous studies have suggested that the functional organization of the brain is hierarchical, analogous decompositions might better capture functional connectivity patterns. Moreover, hierarchical decompositions can efficiently reduce the very high dimensionality of functional connectivity data. This paper provides a novel method for the extraction of hierarchical connectivity components in the human brain using resting-state fMRI. The method builds upon prior work of Sparse Connectivity Patterns (SCPs) by introducing a hierarchy of sparse, potentially overlapping patterns. The components are estimated by cascaded factorization of correlation matrices generated from fMRI. The goal of the paper is to extract sparse interpretable hierarchically-organized patterns using correlation matrices where a low rank decomposition is formed by a linear combination of a higher rank decomposition. We formulate the decomposition as a non-convex optimization problem and solve it using gradient descent algorithms with adaptive step size. Along with the hierarchy, our method aims to capture the heterogeneity of the set of common patterns across individuals. We first validate our model through simulated experiments. We then demonstrate the effectiveness of the developed method on two different real-world datasets by showing that multi-scale hierarchical SCPs are reproducible between sub-samples and are more reproducible as compared to single scale patterns. We also compare our method with an existing hierarchical community detection approach.
\end{abstract} 


\section{Introduction}


It has been known that the human brain consists of spatially different regions which are functionally connected to form networks \cite{sporns2010networks}. In addition, these networks are thought to be hierarchically organized in the brain \cite{doucet2011brain,park2013structural,ferrarini2009hierarchical,meunier2009hierarchical}. However, our understanding of the hierarchical nature of these networks is limited due to their complex nature. Most of the commonly used methods, such as Independent Component Analysis (ICA) \cite{beckmann2005investigations}, Sparse Dictionary Learning (DL) \cite{eavani2012sparse} and graph theory based network analysis \cite{bullmore2009complex}, for analysis of functional networks are focused on estimating a fixed number of networks with no hierarchy. If the assumption about the hierarchy is true then the original data might contain complex hierarchical information with implicit lower-level hidden attributes, that classical one level connectivity methodologies would not be able to capture effectively and interpretably. Notably, existing methods extract different number of components but can not describe relationships between the components.

Most of the methods used for estimation of hierarchical networks in fMRI data analysis are of agglomerative (``bottom-up”) type such as Hierarchical clustering \cite{wang2013analysis,liu2012correlation}, Hierarchical Community Detection \cite{ashourvan2017multi} where the method begins by regarding each element as a separate network and then merging them into larger networks successively. Most of the hierarchical community detection approaches assume that the communities are independent \cite{betzel2017multi, puxeddu2020modular, betzel2015functional} where they have investigated multi-scale brain networks and conducted multi-scale community detection by manipulating the number of communities. But, this is not the case in the human brain where it is known the certain brain regions interact with multiple networks i.e., the networks overlap \cite{xu2016large}. Relatively few hierarchical community detection methods \cite{lancichinetti2009detecting,shen2009detect,lancichinetti2011finding,zhang2015mining} have been developed which find overlapping communities. Moreover, in community detection approaches, negative edge links are treated as repulsion. Previously, most approaches have used thresholds before their analysis and estimated networks by using sparse graphs. The reason for thresholding was that the strong edges contain most relevant information leading to the removal of negative edges. In contrast, in resting fMRI, a negative edge link carries essential information on functional co-variation with the opposing phase \cite{rubinov2011weight} and has a substantial physiological basis \cite{zhan2017significance,fox2005human}. These relations may play an important role in neuropsychiatric disorders and cognitive differentiation \cite{fitzpatrick2007associations}. Some studies have recently shown that the weak network edges contain unique information that can not be revealed by analysis of just strong edges \cite{goulas2015strength,santarnecchi2014efficiency}. Assigning anti-correlated and correlated regions to the same component can reveal more details about the organization of the human brain patterns \cite{eavani2015identifying,sahoo2019sparse,sahoo2018gpu}, as long as interpreted correctly. 

Non-negative Matrix Factorization (NMF) \cite{yang2013overlapping} is one common matrix decomposition approach which many researchers use for obtaining information about community structure by analyzing low dimensional matrix. NMF has been used to find hierarchical structure \cite{song2013hierarchical}, recently, \cite{li2018identification} used Deep Semi Non-negative Matrix Factorization \cite{trigeorgis2017deep} for estimating hierarchical, potentially overlapping, functional networks. The model given by \cite{li2018identification} could only find networks containing regions with positive correlation between them as the method is based on non-negative matrix factorization thus limiting the model to only use positive matrices. 

Our work addresses aforementioned limitations by modeling the fMRI data to capture essential properties of the network, namely- 1) Sparsity: only a small subset of nodes interact with other nodes in a given network; 2) Heterogeneity: some networks might be more prominent in particular individuals as compared to others; 3) Existence of positively and negatively correlated nodes in a network; 4) Overlapping networks, which is likely to reflect true brain organization, as brain networks might share certain regional components; and 5) Hierarchy: By adding extra layers of abstraction we can learn latent attributes and the hierarchy in the networks. Our method is built upon Sparse Connectivity Patterns (SCPs) \cite{eavani2015identifying} which can be considered a symmetric CP decomposition for which an indirect fitting procedure makes the model structure equivalent to the PARAFAC2 model representation considered in \cite{madsen2017quantifying} with the addition of sparsity rather than orthogonality. Our method aims to find Hierarchical Sparse Connectivity Patterns (hSCPs) by jointly decomposing correlation matrices into multiple components having different ranks using a cascaded framework for matrix factorization. We use gradient descent with adaptive step size for solving non-convex optimization, and have also introduced an initialization algorithm for making algorithm deterministic and faster. We evaluate the representation learned by the model on two different real datasets and compare it with EAGLE \cite{shen2009detect} and OSLOM \cite{lancichinetti2011finding} which are well known hierarchical overlapping community detection algorithms. We also provide an extension of the model for clustering the data using hSCPs which could help in understanding of hSCPs and its distribution.

 The organization of the remainder of the paper as follows. In Section 2, we present the method for the extraction of hSCPs shared between rs-fMRI scans. Section 3 presents experimental results for validation of the method on simulated datasets and the effectiveness on the rs-fMRI scans of the 100 unrelated HCP subjects \cite{van2013wu} and 969 subjects from the Philadelphia Neurodevelopmental Cohort (PNC) data set \cite{satterthwaite2014neuroimaging}. We conclude with a discussion.
\section{Method}
\subsection{Sparse Connectivity Patterns}
Let $\mathbf{X}^i \in \mathbb{R}^{P \times T}$ be the fMRI data of the $i^{th}$ subject having $P$ regions and $T$ time points, and $\mathbf{\Theta}^i \in \mathbb{S}^{P \times P}_{++}$ is the correlation matrix where $\mathbf{\Theta}^i_{m,o} = \corr(\mathbf{x}^i_m,\mathbf{x}^i_o)$ is the correlation between time series of $m^{th}$ and $o^{th}$ node. We first define the model for estimating the Sparse Connectivity Patterns (SCPs) \cite{eavani2015identifying} in the fMRI data which decomposes the correlation matrices into non-negative linear combination of sparse low rank components such that for all $i=1,..,S $ we have $\mathbf{\Theta}^i = \mathbf{W} \mathbf{\Lambda}^i \mathbf{W}^T$ where $\mathbf {W} \in \mathbb{R}^{P \times k}$ is a set of shared patterns across all subjects, $k < P$ and $\mathbf{\Lambda}^i \succeq 0$ is a diagonal matrix storing the subject specific information about the strength of each of the components. Let $\mathbf{w}_l \in \mathbb{R}^{P }$ be the $l^{th}$ column of $\mathbf{W}$ such that $-1 \preceq \mathbf{w}_l \preceq 1$ and let $w_{l,s}$ be the $s^{th}$  element of $\mathbf{w}_l$ vector, then $\mathbf{w}_l$ represents a component which reflects the weights of the nodes in the component and if $w_{l,s}$ is zero then $s^{th}$ node does not belong to $l^{th}$ component. If the sign of weights of any two nodes in a component is same then they are positively correlated else they have anti-correlation. To make the patterns sparse, each column of $\mathbf{W}$ was subjected to $L_1$ penalty and the below optimization is solved to obtain the SCPs  
 \begin{equation}
\begin{aligned}
& \underset{\mathbf{W},\mathbf{\Lambda}}{\text{minimize}}
& & \sum_{i=1}^{S}||\mathbf{\Theta}^i - \mathbf{W}\mathbf{\Lambda}^i \mathbf{W}^T ||_F^2\\
& \text{subject to}
&&\|\mathbf{w}_l\|_1 \leq \lambda, l=1,...,k \\
&&&\|\mathbf{w}_l\|_\infty \leq 1, l=1,...,k\\
&&& \mathbf{\Lambda}^i \succeq 0, i=1,...,S
\end{aligned}
   \label{problem}
 \end{equation}
 where $S$ is the total number of  subjects and $\lambda$ controls the sparsity of the components.
 \subsection{Hierarchical Sparse Connectivity Patterns} \label{hSCP}
We have extended the above work and introduced Hierarchical Sparse Connectivity Patterns (hSCPs) to estimate hierarchical sparse low rank patterns in the correlation matrices. In our model, a correlation matrix is decomposed into $K$ levels as - 
\begin{align}
\begin{split}
\mathbf{\Theta}^i & \approx \mathbf{W}_1\mathbf{\Lambda}_1^i \mathbf{W}_1^T \\
\mathbf{\Theta}^i & \approx \mathbf{W}_1\mathbf{W}_2\mathbf{\Lambda}_2^i \mathbf{W}_2^T\mathbf{W}_1^T \\ .  \\\mathbf{\Theta}^i & \approx  \mathbf{W}_1\mathbf{W}_2..\mathbf{W}_K \mathbf{\Lambda}^i_K \mathbf{W}_K^T\mathbf{W}_{K-1}^T..\mathbf{W}_1^T
  \end{split}
\end{align}
where $\mathbf{W}_1 \in  \mathbb{R}^{P \times k_1}$ and $\mathbf{W}_q \in  \mathbb{R}^{k_{q-1} \times k_q}$, $\mathbf{\Lambda}_q^i \in  \mathbb{R}^{k_q \times k_q}$ is a diagonal matrix storing subject specific information of the patterns, $P \gg k_1 > k_2 > ... > k_K $, $P \gg K $ and $\mathbf{W}^T $ is the transpose of $\mathbf{W}$. Here $k_r$ is the number of components at the $r^{th}$ level, note that $k_1$ is the number of components at the lower most level of the hierarchy. If we consider 2 layer hierarchical representation of a given correlation matrix then we can define $\mathbf{Z}_1 = \mathbf{W}_1\mathbf{W}_2$ to be a $P \times k_2$ matrix, then $\mathbf{Z}_1$ is a coarse network which consist of weighted linear combination of $\mathbf{W}_1$ which are fine level components where weights are stored in $\mathbf{W}_2$. 
 
 For better interpretability, for noise reduction in the model, but also because of our hypothesis that brain subnetworks are relatively sparse \cite{achard2007efficiency}, we have introduced sparsity constraints on the $\mathbf{W}$ matrices. By making $\mathbf{W}_1$ sparse we are forcing the components to contain few number of nodes and by forcing rest of the $\mathbf{W}$s to be sparse, we are forcing that the components at each of the next level are sparse linear combination of previous components. The hierarchical networks can be estimated by solving the below minimization procedures simultaneously under the constraints mentioned above
 \begin{align}
\begin{split}
\underset{{W_1},{\Lambda_1}}{\text{min}}& \sum_{i=1}^{S}\|\mathbf{\Theta}^i - \mathbf{W}_1\mathbf{\Lambda}_1^i \mathbf{W}_1^T\|_F^2 \\
\underset{{W_1,W_2},{\Lambda_2}}{\text{min}}&\sum_{i=1}^{S}\|\mathbf{\Theta}^i - \mathbf{W}_1\mathbf{W}_2\mathbf{\Lambda}_2^i \mathbf{W}_2^T\mathbf{W}_1^T \|_F^2\\ . \\\underset{\mathcal{W},{\Lambda_K}}{\text{min}}  &\sum_{i=1}^{S}\| \mathbf{\Theta}^i -  \mathbf{W}_1\mathbf{W}_2..\mathbf{W}_K \mathbf{\Lambda}^i_K \mathbf{W}_K^T\mathbf{W}_{K-1}^T..\mathbf{W}_1^T\|_F^2
  \label{eqn:model}
  \end{split}
\end{align}
 where $\mathcal{W}=\{\mathbf{W}_1,..,\mathbf{W}_K \}$. As the above minimization procedures are inter-dependent, we need to solve them jointly. Let $\mathcal{L}=\{\mathbf{\Lambda}_1,..,\mathbf{\Lambda}_K \}$ and $H(\mathcal{W},\mathcal{L}) = \sum_{i=1}^{S} \sum_{r=1}^{K}||\mathbf{\Theta}^i - (\Pi_{j=1}^{r}\mathbf{W}_j)\mathbf{\Lambda}_r^i(\Pi_{n=1}^{r}\mathbf{W}_n)^T ||_F^2$. The joint minimization problem can be written as below
 \begin{equation}
\begin{aligned}
& \underset{\mathcal{W},\mathcal{L}}{\text{minimize}}
& & H(\mathcal{W},\mathcal{L}) \\
& \text{subject to}
&&\|\mathbf{w}^r_l\|_1 < \lambda_r, l=1,...,k_r \hspace{0.3cm} \text{and} \hspace{0.3cm} r=1,..,K  \\
&&&\|\mathbf{w}^r_l\|_\infty \leq 1, l=1,...,k_r \hspace{0.3cm} \text{and} \hspace{0.3cm} r=1,..,K \\
&&&\mathbf{W}_j \geq 0, j=2,...,K\\
&&& \mathbf{\Lambda}_r^i \succeq 0, i=1,...,S  \hspace{0.3cm} \text{and} \hspace{0.3cm} r=1,..,K \\
&&& \trace(\mathbf{\Lambda}_r^i) =1, i=1,...,S  \hspace{0.3cm} \text{and} \hspace{0.3cm} r=1,..,K 
\end{aligned}
   \label{problem1}
 \end{equation}
 
where $\trace$ operator calculates sum of diagonal elements of a matrix. In the above minimization procedure, the sum of diagonal values of $\Lambda^i$ is fixed to be $1$ such that the sparsity of $W$ is not trivially minimized. The optimization problem defined in \ref{problem1} is a non-convex problem which we solved using alternating minimization. Below are the gradients of $H$ with respect to $\mathcal{W}$ and $\mathcal{L}$. Let us first define the following variables
\begin{align*}
&\mathbf{W}_0 = \mathbf{I}_P\\
&\mathbf{Y}_r = \Pi_{j=0}^{r}\mathbf{W}_j \\
&\mathbf{T}_{n,i}^r = (\Pi_{j=1}^{n-r}\mathbf{W}_j)\mathbf{\Lambda}^i_{n-r}(\Pi_{j=1}^{n-r}\mathbf{W}_j)^T
\end{align*}
the gradient of $H$ with respect to $\mathbf{\Lambda}_r^i$ is:
\begin{equation}
\begin{aligned}
\frac{\partial H}{\partial \mathbf{\Lambda}_r^i} &= (-2\mathbf{Y}_r^T\mathbf{\Theta}_r^i\mathbf{Y}_r + 2\mathbf{Y}_r^T\mathbf{Y}_r\mathbf{\Lambda}_r^i\mathbf{Y}_r^T\mathbf{Y}_r) \circ \mathbf{I}_{k_r}
\end{aligned}
\label{grad1}
\end{equation}
where $\circ$ is entry-wise product. The gradient of $H$ with respect to $\mathbf{W}_l$ is written as:
\begin{equation}
\begin{aligned}
\frac{\partial H}{\partial \mathbf{W}_r} &= \sum_{i=1}^S\sum_{j=r}^{K} -4\mathbf{Y}_{r-1}^T\mathbf{\Theta}_i\mathbf{Y}_{r-1}\mathbf{W}_r\mathbf{T}_{j,i}^r \\ & +   4\mathbf{Y}_{r-1}^T\mathbf{Y}_{r-1}\mathbf{W}_r\mathbf{T}_{j,i}^r\mathbf{W}_r^T\mathbf{Y}_{r-1}^T\mathbf{Y}_{r-1}\mathbf{W}_r\mathbf{T}_{j,i}^r  
\end{aligned}
\label{grad2}
\end{equation}
Algorithm~\ref{alg:HSCP} describes the complete alternating minimization procedure where $\proj_1(\mathbf{W},\lambda)$ operator projects each column of $\mathbf{W}$ into intersection of $L_1$ and $L_\infty$ ball \cite{podosinnikova2013robust}, and $\proj_2$ projects a matrix onto $\mathbb{R}_{+}$ by making all the negative elements in the matrix equal to zero. As the gradients are not globally Lipschitzs, we don't have bounds on the step size for the gradients. For that reason, we have used AMSGrad \cite{reddi2019convergence}, ADAM \cite{kingma2014adam} and NADAM \cite{dozat2016incorporating} as gradient descent algorithms which have adaptive step size. $\descent$ function in the Algorithm~\ref{alg:HSCP} is the update rule used by different gradient descent techniques. All the code is implemented in \textsc{matlab} and will be released upon publication. The cost of computing gradients of $\mathbf{\Lambda}$ is $\bigO(KSP^2k_1)$ and of $\mathbf{W}$ is $\bigO(KSP^2k_1 + K^2SPk_1^2)$. The overall cost of Algorithm 1 is number of iterations $\times \bigO(KSP^2k_1 + K^2SPk_1^2) $. From our previous assumption that $P \gg K $, the final cost is number of iterations $\times \bigO(KSP^2k_1 ) $.

 In the above formulation, the last level has the highest number of components $k_1$, and in the level after that we have $k_2$ number of components which are linear combination of components at previous level, so on and so forth. In this way, we have built up a hierarchical model where each component is made up of linear combination of components at the previous hierarchy. Note that we can not just use the last decomposition in the above architecture to get the hierarchy as different layers have different ranks and different approximations, hence we will need all the approximations to build the hierarchical structure. In addition, one would expect $\mathbf{W}_2$ and $\mathbf{W}$s to be degenerate, but that would be the case only when $\mathbf{W}_1$ is orthogonal matrix. Consider the case where we have a two level hierarchy, we can have better approximation by taking a linear combination of columns of $\textbf{W}_1$ which we have also observed empirically.

 \begin{algorithm}[t]
 \caption{hSCP}\label{alg:HSCP}
\begin{algorithmic}[1]
\State \textbf{Input:} Data $\Theta$, number of connectivity patterns $k_1$,..,$k_K$ and sparsity $\lambda_1$,..,$\lambda_K$ at different level 
\State $\mathcal{W}$ and $\mathcal{L}$ = Initialization($\Theta$) 
\Repeat 
\For{$r=1$ {\bfseries to} $K$} 
\State $\mathbf{W}_r \leftarrow \descent(\mathbf{W}_r)$
\If{$r==1$}
   \State $\mathbf{W}_r \leftarrow \proj_1(\mathbf{W}_r,\lambda_r) $
   \Else{}
\State $\mathbf{W}_r \leftarrow \proj_2(\mathbf{W}_r) $
   \EndIf
\For{$i= 1,..,S$} 
\State $\mathbf{\Lambda}_r^i \leftarrow \descent(\mathbf{\Lambda}_r^i)$
\State $\mathbf{\Lambda}_r^i \leftarrow \proj_2(\mathbf{\Lambda}_r^i) $
\EndFor
\EndFor
\Until{Stopping criterion is reached}
\State \textbf{Output:} $\mathcal{W}$ and $\mathcal{L}$
\end{algorithmic}
\end{algorithm}

 \begin{algorithm}[H]
 \caption{Initialization}\label{alg:initial}
\begin{algorithmic}[1]
   \State {\bfseries Input:} Data $\Theta$
\For{$r=1$ {\bfseries to} $K$}
   \For{ $i=1$ {\bfseries to} $S$}
   \If{$r==1$}
   \State $ \mathbf{U}^i\mathbf{V}^i(\mathbf{U}^i)^T= k_1\text{- rank SVD}(\mathbf{\Theta}^i)$
   \Else{}
      \State $\mathbf{V}^i = k_{r-1}\text{top values of } \mathbf{\Lambda}_{r-1}^{i}$
       \State $\mathbf{U}^i = \text{Permutation matrix}$
   \EndIf
      \State $\mathbf{\Lambda}_r^i = \mathbf{V}^i$ 
   \EndFor
   \State $\mathbf{W}_r = \frac{1}{S}\sum_{i=1}^{S}(\mathbf{U}^i)$
   \EndFor
   \State {\bfseries Output:} $\mathcal{W}$ and $\mathcal{L}$
\end{algorithmic}
\end{algorithm}

\begin{figure*}
    \begin{minipage}[t]{.49\textwidth}
        \centering
        \includegraphics[trim={2cm 0.1cm 3cm 0.1cm},clip,width=\textwidth]{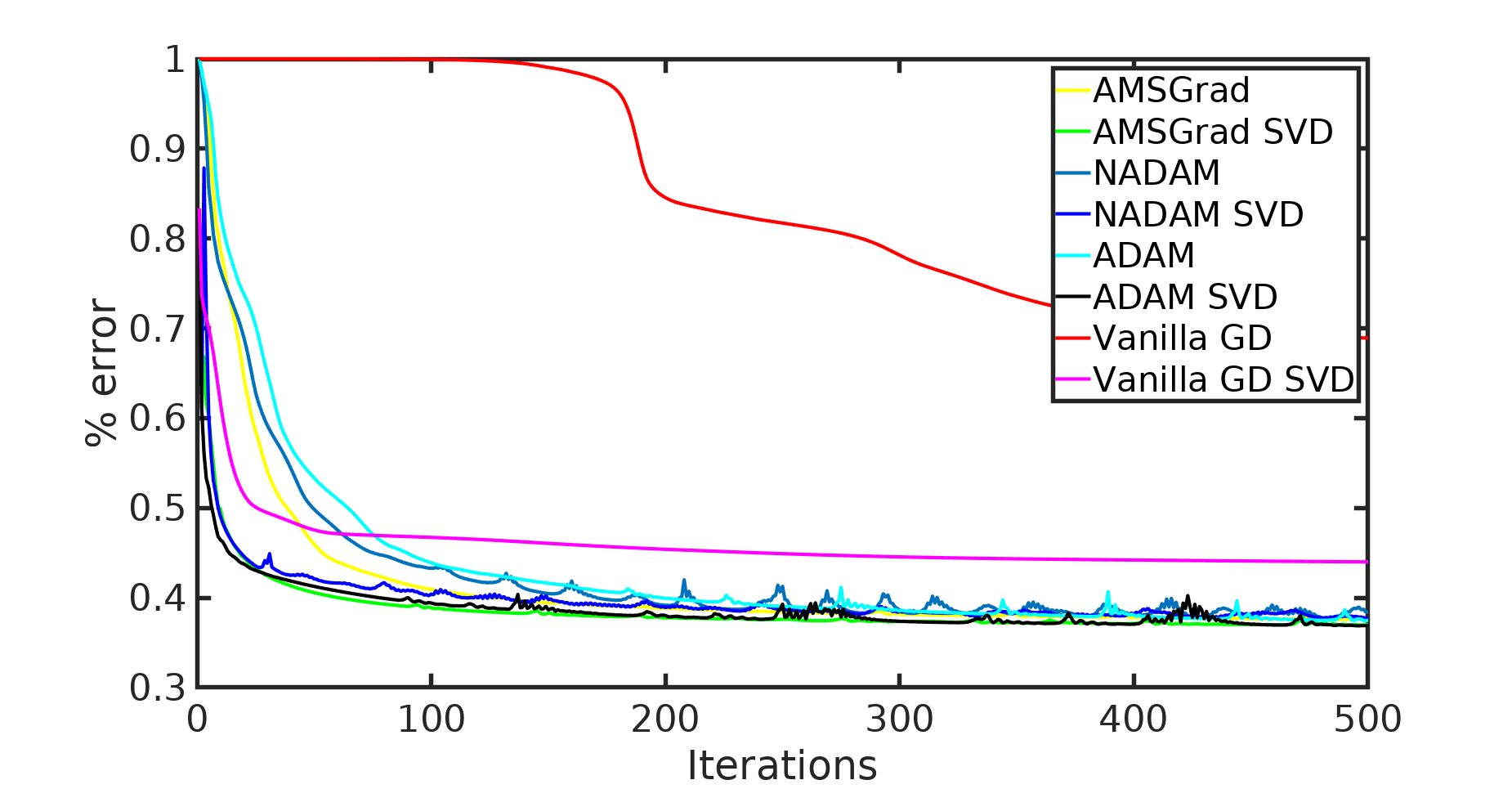}
        \subcaption{$k_1 = 15, k_2 = 5, \lambda_1 = 9, \lambda_2 = 7.5$}
    \end{minipage}
    \begin{minipage}[t]{.49\textwidth}
        \centering
        \includegraphics[trim={2cm 0.1cm 3cm 0.1cm},clip,width=\textwidth]{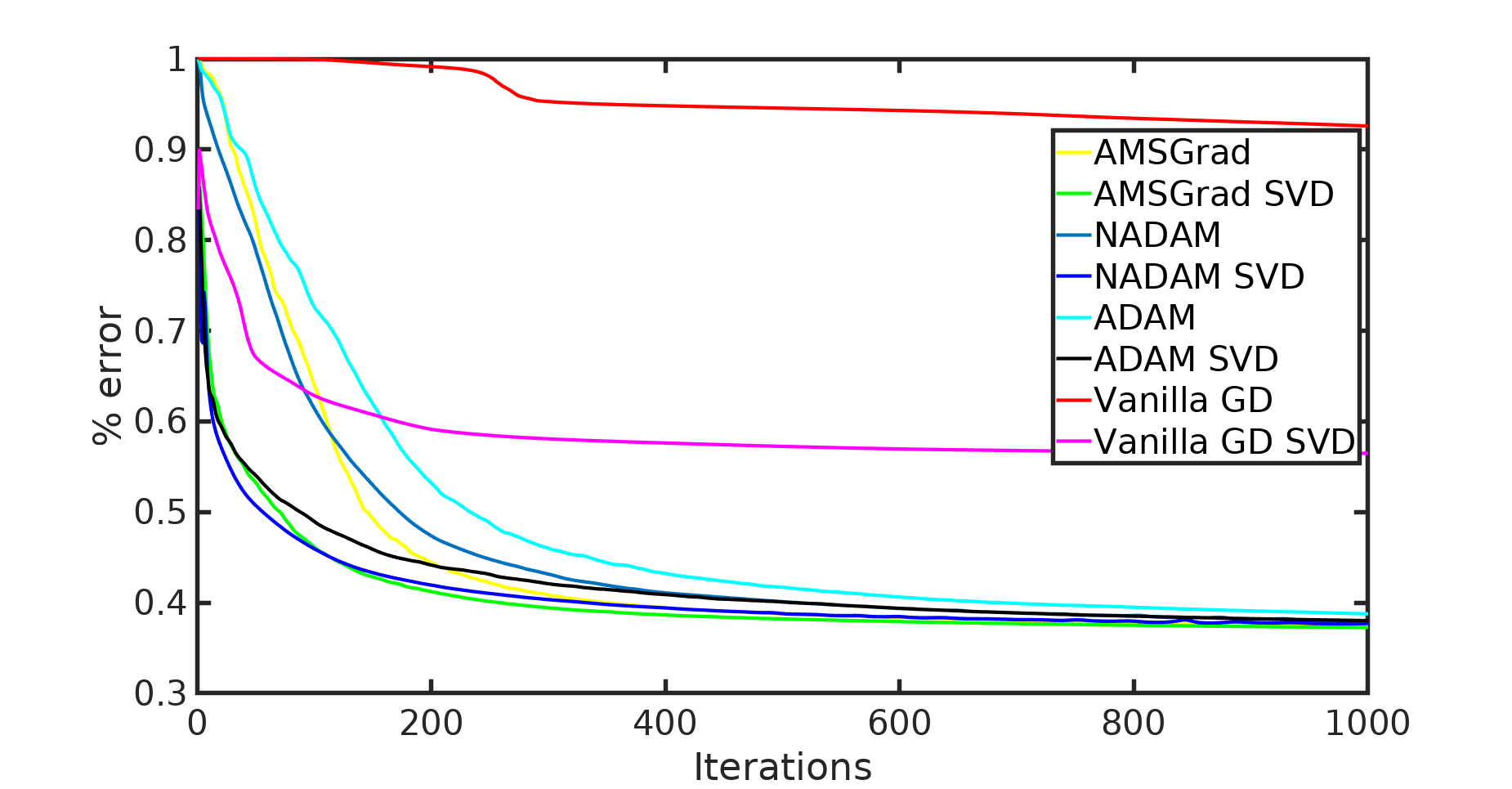}
        \subcaption{$k_1 = 15, k_2 = 5, \lambda_1 = 3.6, \lambda_2 = 3$}
    \end{minipage}

    \caption{Performance comparison of different gradient descent techniques. SVD corresponds to gradient descent with SVD initialization }
    \label{fig:gd}
\end{figure*}
 \subsection{Initialization procedure for Gradient Descent}
Single level matrix decomposition considered in hSCP is structurally similar to Singular Value Decomposition (SVD) but with the dependent components and sparsity added. Hence, we believe that the final components estimated are a modification of singular vectors. Thus, we have initialized the $\mathcal{W}$ and $\mathcal{L}$ in Algorithm $1$ by taking SVD of input data matrix. This helps in making algorithm deterministic. Define $\Bar{\Theta}$ as the sample mean of $\Theta_i$. We then do k-rank SVD of $\Bar{\Theta}$ and obtain $U$ and $S$ such that $UVU^T =$ k-rank SVD of $\Bar{\Theta}$. We then initialize $W_1$ by $U$ and $\Lambda_1^i$ by $V^i$ where $V^i$ can be obtained by taking k-rank SVD of $\Theta_i$ as described in Algorithm~\ref{alg:initial}. For $r > 1$, $W_r$ can be initialized as a permutation matrix and $\Lambda_r$ by top $k_r$ diagonal elements of $k_{r-1}$ so that we don't have to perform SVD at each level. We empirically show in the next section that SVD initialization results in faster convergence. 

\begin{figure*}
    \begin{minipage}[t]{.49\textwidth}
        \centering
        \includegraphics[trim={2cm 0.1cm 3cm 0.1cm},clip,width=\textwidth]{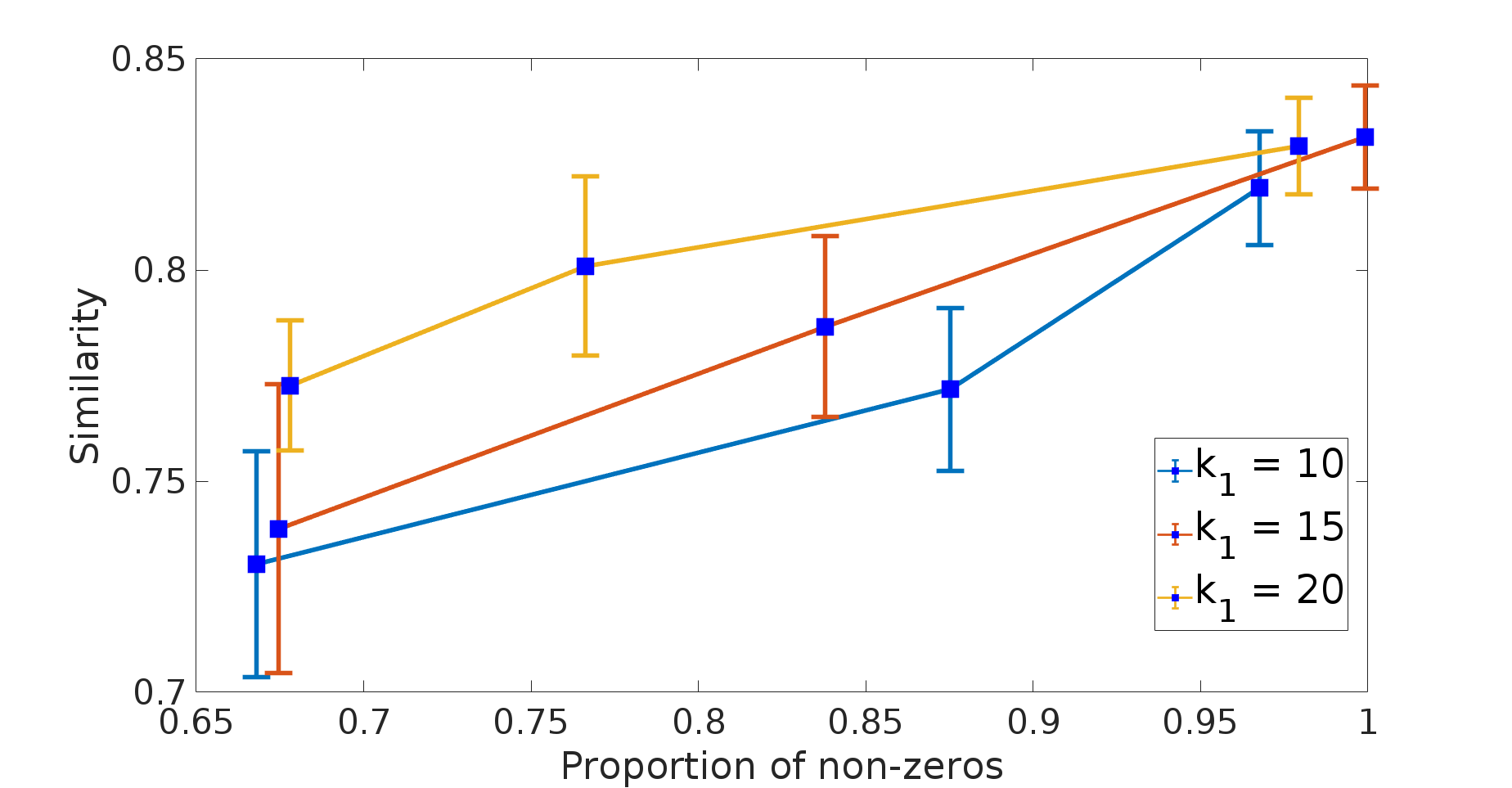}
        \subcaption{Similarity comparison at fine scale}
    \end{minipage}
    \begin{minipage}[t]{.49\textwidth}
        \centering
        \includegraphics[trim={2cm 0.1cm 3cm 0.1cm},clip,width=\textwidth]{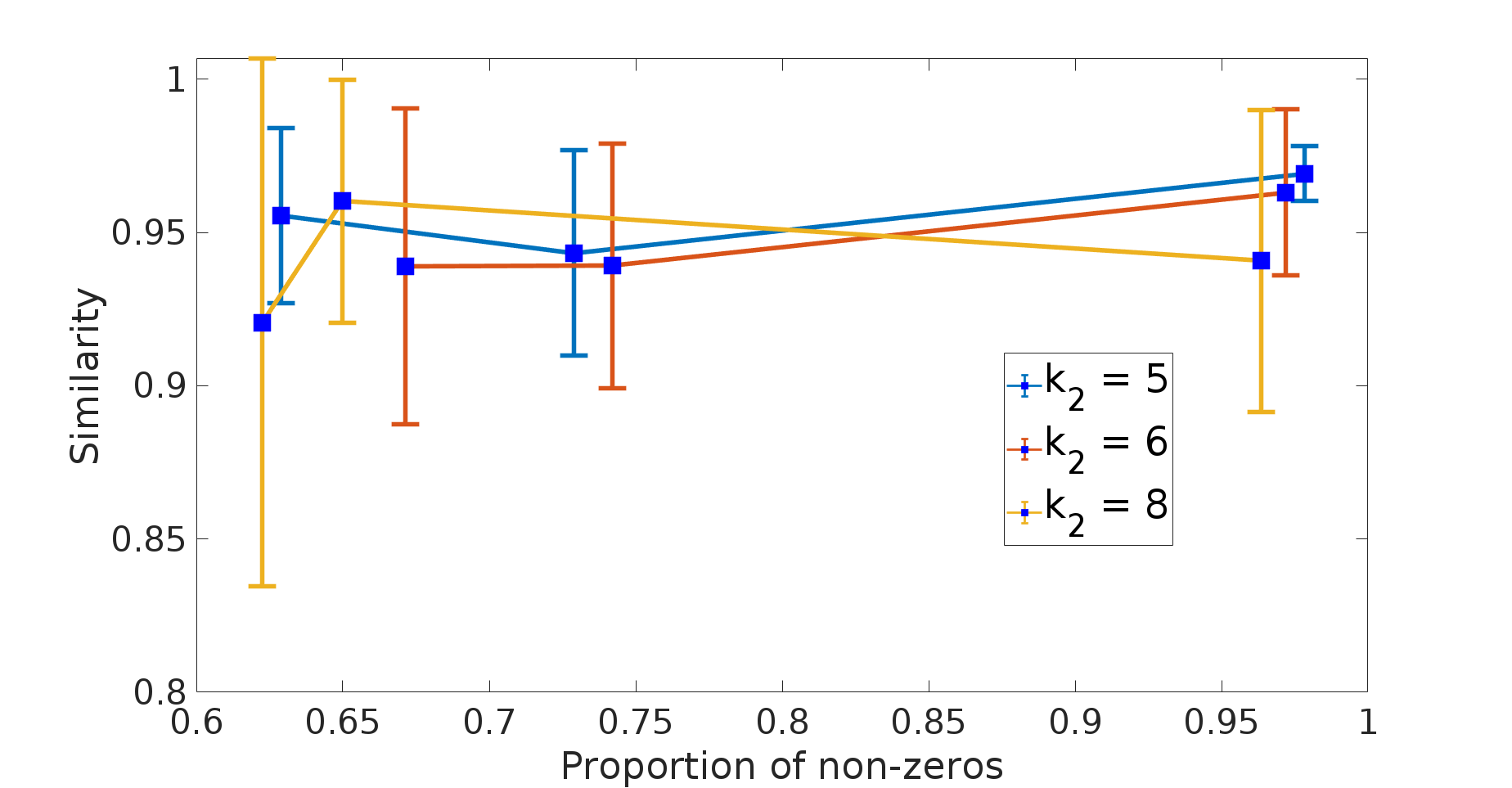}
        \subcaption{Similarity comparison at coarse scale}
    \end{minipage}

    \caption{Comparison between ground truth and extracted hSCPS components on simulated dataset. X axis corresponds to proportion of non-zeros in the estimated components. }
    \label{fig:simu}
\end{figure*}

\section{Experiments}

\subsection{Dataset}
We used two real dataset for demonstrating the effectiveness of the method which are described below 
\begin{itemize}[leftmargin=*]
    \item HCP- Human Connectome Project (HCP) \cite{van2013wu} dataset is one of the widely used dataset for fMRI analysis containing fMRI scans of $100$ unrelated subjects as provided at the HCP $900$ subjects data release \cite{van2012human} which were processed using ICA+FIX pipeline with MSMAll registration \cite{glasser2013minimal}. {Each subject has $4004$ time points and the time series were normalized to zero mean and unit L2 norm, averaged over the 360 nodes of the multimodal HCP parcellation \cite{glasser2016multi}.}
\item  PNC- Philadelphia Neuro-developmental Cohort (PNC) \cite{satterthwaite2014neuroimaging} dataset contains $969$ subjects (ages from $8$ to $22$) each having $120$ time points and $121$ nodes described in \cite{doshi2016muse}. The data were preprocessed using an optimized procedure \cite{ciric2017benchmarking} which includes slice timing, confound regression, and band-pass filtering.
\end{itemize}
\subsection{Convergence Analysis}

We compare AMSGrad, ADAM, NADAM and vanilla gradient descent with SVD initialization and random initialization by measuring percentage error which is defined as:
\begin{align*}
    \frac{\sum_{i=1}^{S} \sum_{r=1}^{K}||\mathbf{\Theta}^i - (\Pi_{j=1}^{r}\mathbf{W}_j)\mathbf{\Lambda}_r^i(\Pi_{n=1}^{r}\mathbf{W}_n)^T ||_F^2}{\sum_{i=1}^{S} \sum_{r=1}^{K}||\mathbf{\Theta}^i ||_F^2}
\end{align*}
For fair comparison, we set $\beta_1 = 0.9$ and $\beta_2 = 0.99$ for ADAM, NADAM and AMSGRAD algorithm, where $\beta_1$ and  $\beta_2$ are the hyperparameters used in the update rules of the gradient descent algorithms. These are values are typically used as parameter settings for adaptive gradient descent algorithms \cite{reddi2019convergence}. Figure~\ref{fig:gd} shows the convergence of the algorithm on the complete HCP data for two different combinations of sparsity parameters at a particular set of $k_1$ and $k_2$. From the Figure~\ref{fig:gd} we can see that the AMSGrad has the best convergence and SVD initialization gives a better convergence rate. For rest of the experiments we have used AMSGrad algorithm with SVD initialization to perform gradient descent.

\begin{table}
  \begin{tabularx}{0.48\textwidth}{p{.07cm}p{0.7cm}p{2cm}p{2cm}p{2cm}}
    \toprule
    
    & & {$10$}   & {$15$} & {$20$} \\
    \midrule
        \multirow{ 3}{*}{$5$} & hSCP & $0.8293\pm 0.0467$ &  $0.8097\pm 0.0728$ &   $0.8305\pm 0.0614$   \\   
    & EAGLE & $0.4051\pm 0.0304$  & $0.4180\pm 0.0290$ & $0.4068\pm 0.0070$      \\
    & OSLOM & $0.6866\pm 0.0442$ & $0.6955\pm 0.0362$ &-      \\
    \midrule
        \multirow{ 3}{*}{$6$}& hSCP & $0.8421\pm 0.0585$ &  $0.8660\pm 0.0286$ &   $0.8497\pm 0.0292$ \\
   &EAGLE& $0.3867\pm 0.0141$  & $0.4855\pm 0.0731$ & $0.4463\pm 0.0334$       \\
   & OSLOM  & $0.6249\pm 0.0554$  & $0.7302\pm 0.0431$ & -     \\
   \midrule
       \multirow{ 3}{*}{$8$}& hSCP & $0.8350\pm 0.0666$ &    $0.8457\pm 0.0353$ &    $0.8454 \pm 0.0385$    \\
   &EAGLE & $0.4408\pm 0.0857$  & $0.5339 \pm 0.0900$ &  $0.4099 \pm 0.0274$     \\
    &OSLOM & $0.6610 \pm 0.0540$  & - &  -      \\
    
    \bottomrule
  \end{tabularx}\\
  \captionof{table}{{Similarity comparison (mean$\pm$std) on simulated dataset. The rows correspond to  values of $k_1$ and the columns correspond to values of $k_2$.}}
  \label{table:simu}
\end{table}

\subsection{Simulation} 
To evaluate the performance of the proposed model, we first use synthetic data. We compared the hierarchical components extracted from hSCP to hierarchical overlapping communities obtained using EAGLE \cite{shen2009detect} and OSLOM \cite{lancichinetti2011finding}. Implementation of EAGLE and OSLOM was obtained from the authors. 

We randomly generate $\mathbf{V}_1 \in \mathbb{R}^{p \times k_1}$ with percentage of non-zeros equal to $\mu_1$, $\mathbf{W}_2 \in \mathbb{R}^{k_1 \times k_2}$ with percentage of non-zeros equal to $\mu_2$ and $\mathbf{\Lambda}^i \in \mathbb{R}^{k_2 \times k_2}$ for $i = 1, ..,n$. The goal is to generate $\mathbf{V}_1\mathbf{W}_2\mathbf{\Lambda}^i \mathbf{W}_2^T\mathbf{V}_1^T$ matrices which are close to a correlation matrix. For this, we first take mean of  all $\mathbf{\Lambda}^i$ such that $\mathbf{U} = \frac{1}{n}\sum_{i=1}^n \mathbf{\Lambda}^i$ and generate $\mathbf{T}$ such that $\mathbf{T} = \mathbf{V}_1\mathbf{W}_2\mathbf{U} \mathbf{W}_2^T\mathbf{V}_1^T$. Now, let $\mathbf{D}$ be a matrix containing diagonal elements of $\mathbf{T}$, to make $\mathbf{T}$ a correlation matrix, we modify $\mathbf{V}_1$ by multiplying it by $\mathbf{D}^{\frac{1}{2}}$. Let $\mathbf{W}_1 =\mathbf{D}^{\frac{1}{2}}\mathbf{V}_1 $, then $\mathbf{R} = \mathbf{W}_1\mathbf{W}_2\mathbf{U} \mathbf{W}_2^T\mathbf{W}_1^T$ would be a correlation matrix. Now, we generate correlation matrix for each subject by using the below equation
\begin{align*}
   \mathbf{\Theta}^i = \mathbf{W}_1\mathbf{W}_2\mathbf{\Lambda}^i \mathbf{W}_2^T\mathbf{W}_1^T + \mathbf{E}_i
\end{align*}
As $\mathbf{W}_1\mathbf{W}_2\mathbf{\Lambda}^i \mathbf{W}_2^T\mathbf{W}_1^T $ matrix is close to a correlation but not a correlation matrix, we add $\mathbf{E}_i \in \mathbb{R}^{p \times p}$ such that it becomes a correlation matrix. For the experiments, the parameters were set as follows: $n = 300$, $p = 100$ $k_1 = 20$, $k_2 = 10$, $\mu_1 = 0.4$ and $\mu_2 = 0.5$. 

We compare components derived from hSCP with $k_1 \in \{10,15,20\}$, $k_2\in\{5,6,8\}$, $\lambda_1 \in P\times 5(10^{[-3:-1]})$ and $\lambda_2 \in k_1\times10^{[-3:-1]}$. By varying $\lambda$ values, we generate components with different sparsity. We first compare fine-scale and coarse-scale components separately to demonstrate the effect of sparsity on the performance. For a fixed $k_1$ and $\lambda_1$, we find $k_2$ and $\lambda_2$ giving the maximum similarity with the ground truth and for a fixed $k_2$ and $\lambda_2$, we find $k_1$ and $\lambda_1$ giving the maximum similarity with the ground truth over $10$ runs. Here the similarity is defined as the average correlation between extracted and the ground truth components. Fig. \ref{fig:simu} shows the similarity of the fine-scale and coarse-scale components with the ground truth. From the figure, we can see that the hSCP can extract components that are highly similar to the ground truth. Also, as the fine-scale components become sparse, the similarity decreases. Next, we compare hSCP to EAGLE and OSLOM. Hierarchical components and communities with $k_1 \in \{5,6,8\}$, $k_2\in\{10,15,20\}$ were extracted from hSCP, EAGLE and OSLOM. The correlation matrix averaged across all the subjects was used as an input to EAGLE and OSLOM. For hSCP, among different values of $\lambda_1$ and $\lambda_2$, we extract components at level $k_1$ and $k_2$, which have maximum similarity with the ground truth. Table \ref{table:simu} shows the similarity of the extracted components with the ground truth. Some cells in the tables are empty as the EAGLE and OSLOM algorithms were not able to generate hierarchical structures for particular values of $k_1$ and $k_2$. It can be seen that the hSCP method can extract the components which are closer to ground truth as compared to other methods.


        \begin{figure*}[h!btp]
    
       \begin{minipage}[t]{0.49\textwidth}
        \centering
                \includegraphics[trim={0cm 0.1cm 4cm 0.1cm},clip,trim={0cm 0.1cm 3cm 0.1cm},clip,width=0.8\textwidth]{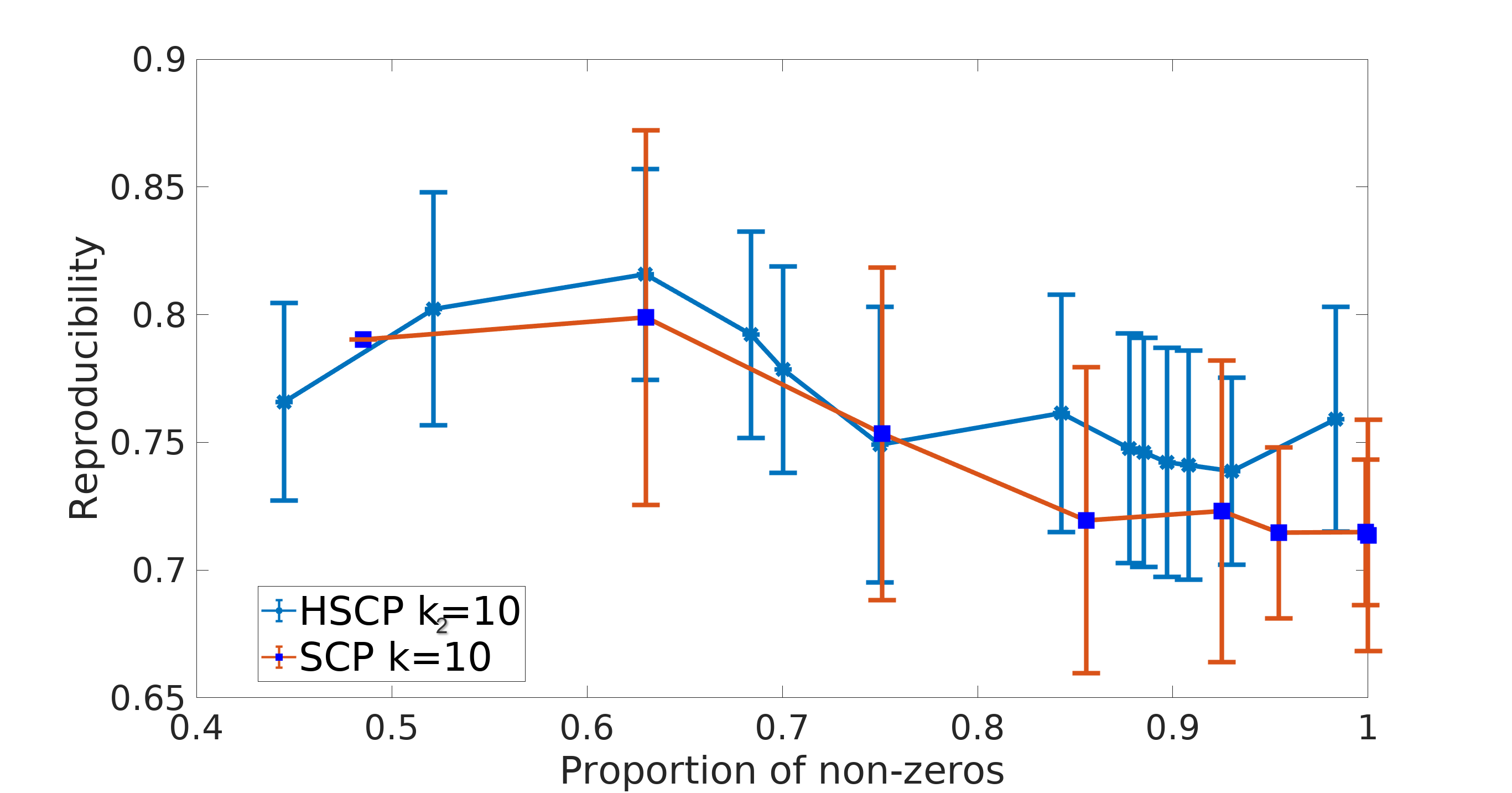}

    \end{minipage}
    \hfill
       \begin{minipage}[t]{0.49\textwidth}
        \centering
                \includegraphics[trim={0cm 0.1cm 4cm 0.1cm},clip,trim={0cm 0.1cm 3cm 0.1cm},clip,width=0.8\textwidth]{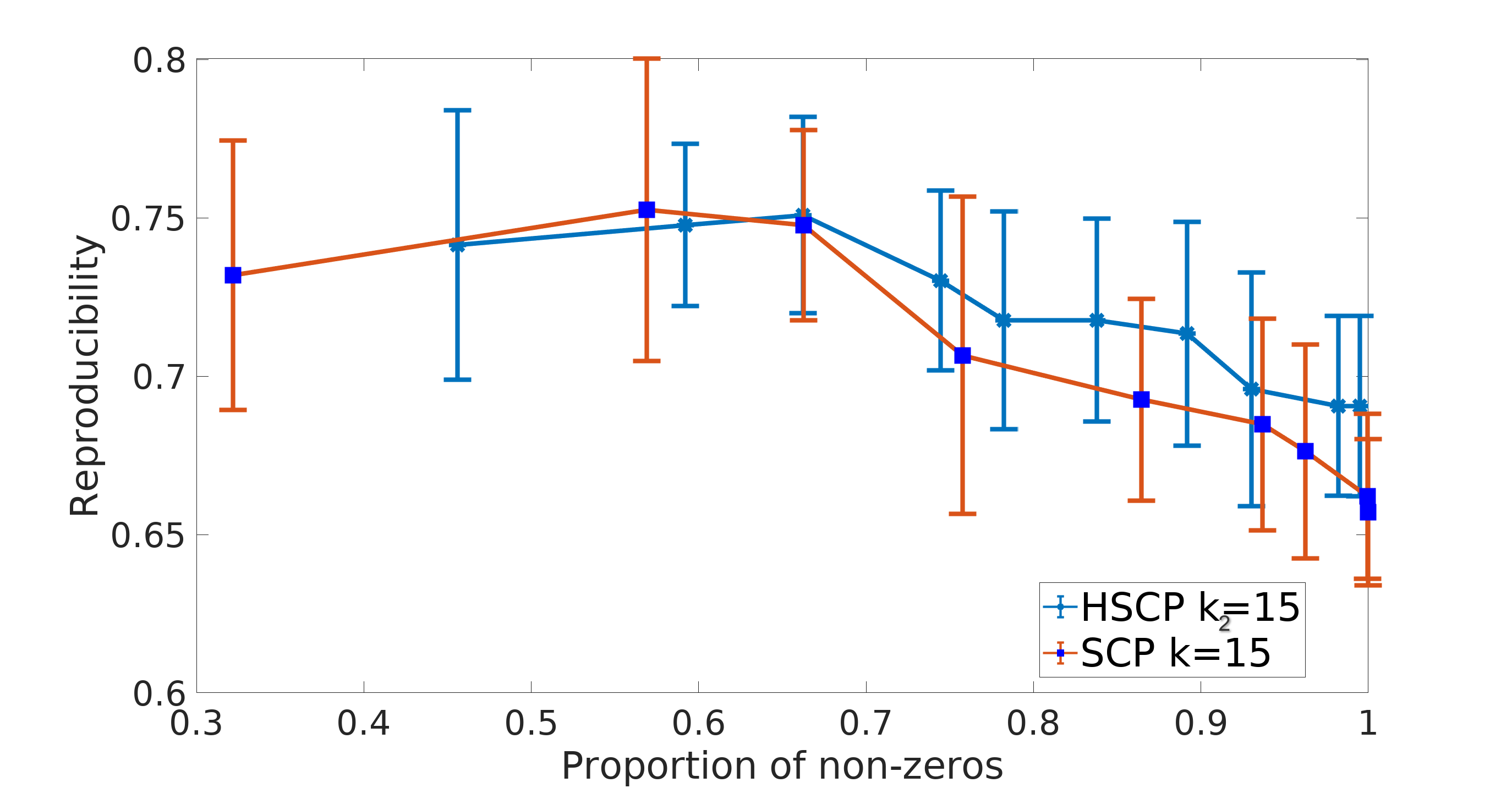}

    \end{minipage}
    \begin{minipage}[t]{0.49\textwidth}
        \centering
                \includegraphics[trim={0cm 0.1cm 4cm 0.1cm},clip,width=0.8\textwidth]{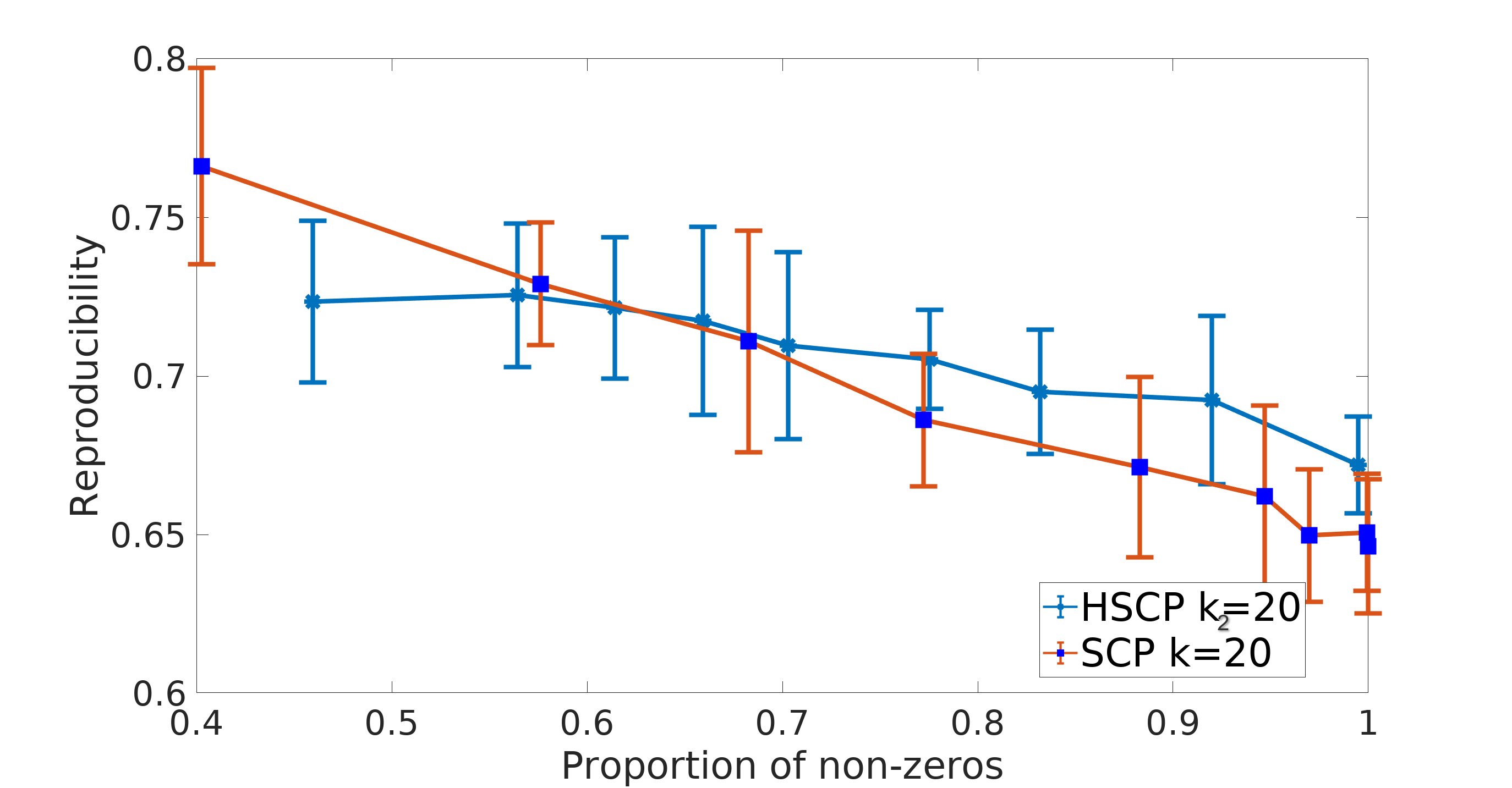}

    \end{minipage}
    \hfill
    \begin{minipage}[t]{0.49\textwidth}
        \centering
                \includegraphics[trim={0cm 0.1cm 4cm 0.1cm},clip,width=0.8\textwidth]{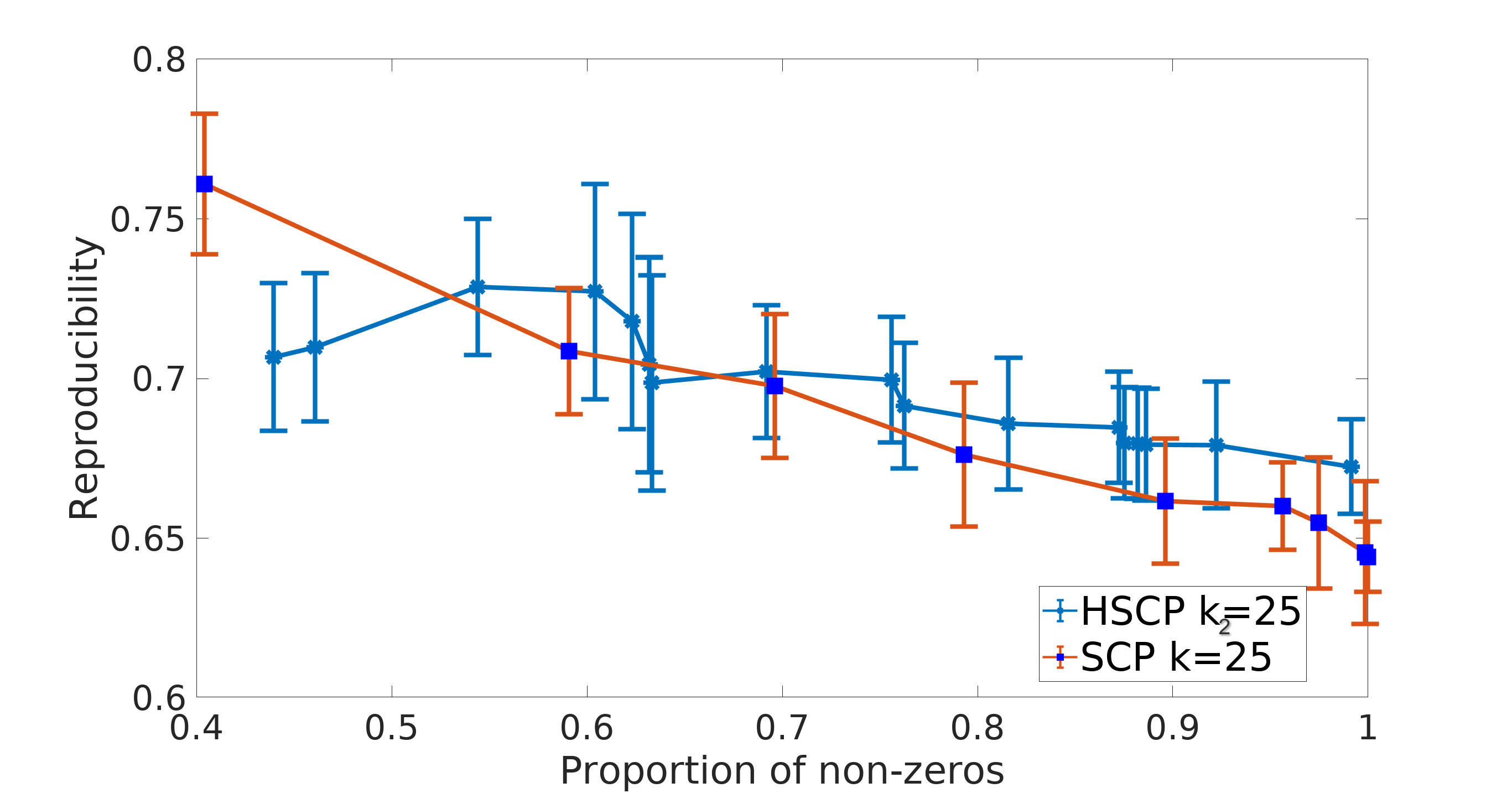}

    \end{minipage}

        \caption{Comparison of single scale (SCP) and hierarchical (HSCP) components on HCP dataset. X axis corresponds to proportion of non-zeros in the components. All the HSCP are second level components.}
    \label{fig:single_hcp}
    
\end{figure*}

        \begin{figure*}[th!bp]
    
       \begin{minipage}[t]{0.49\textwidth}
        \centering
                \includegraphics[trim={0cm 0.1cm 0cm 0.1cm},clip,width=0.8\textwidth]{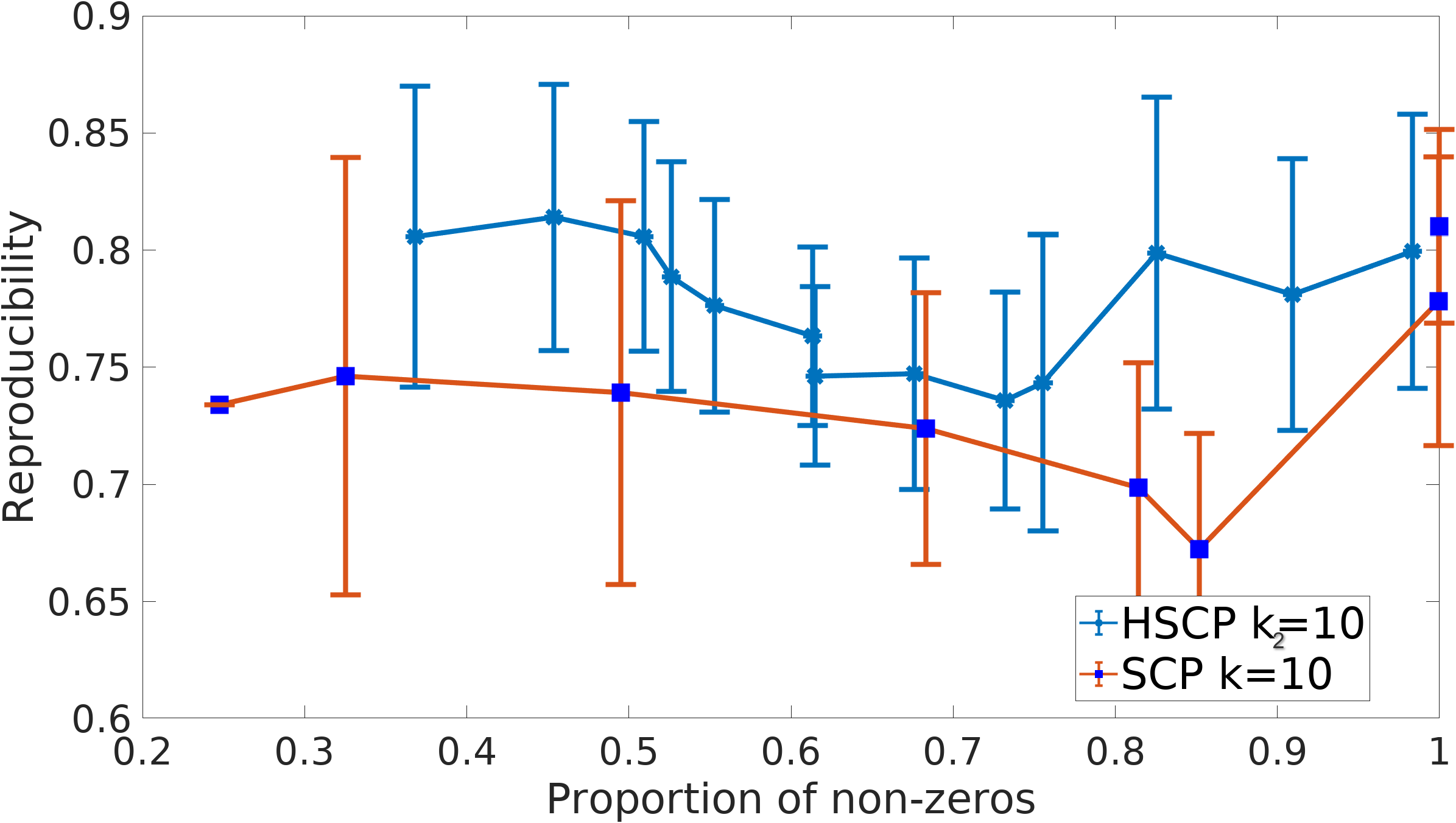}

    \end{minipage}
    \hfill
       \begin{minipage}[t]{0.49\textwidth}
        \centering
                \includegraphics[trim={0cm 0.1cm 0cm 0.1cm},clip,width=0.8\textwidth]{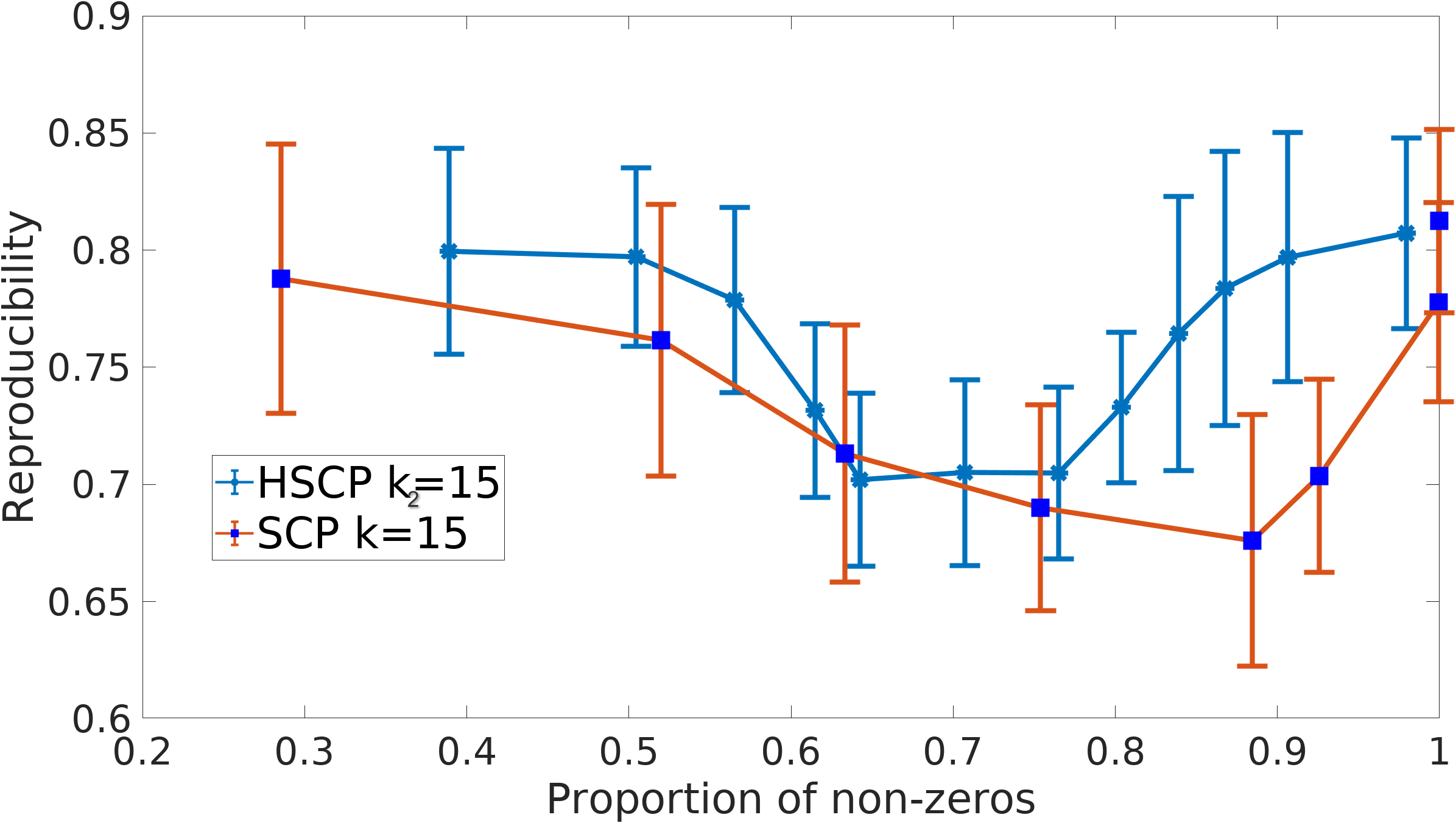}

    \end{minipage}
    \begin{minipage}[t]{0.49\textwidth}
        \centering
                \includegraphics[trim={0cm 0.1cm 0cm 0.1cm},clip,width=0.8\textwidth]{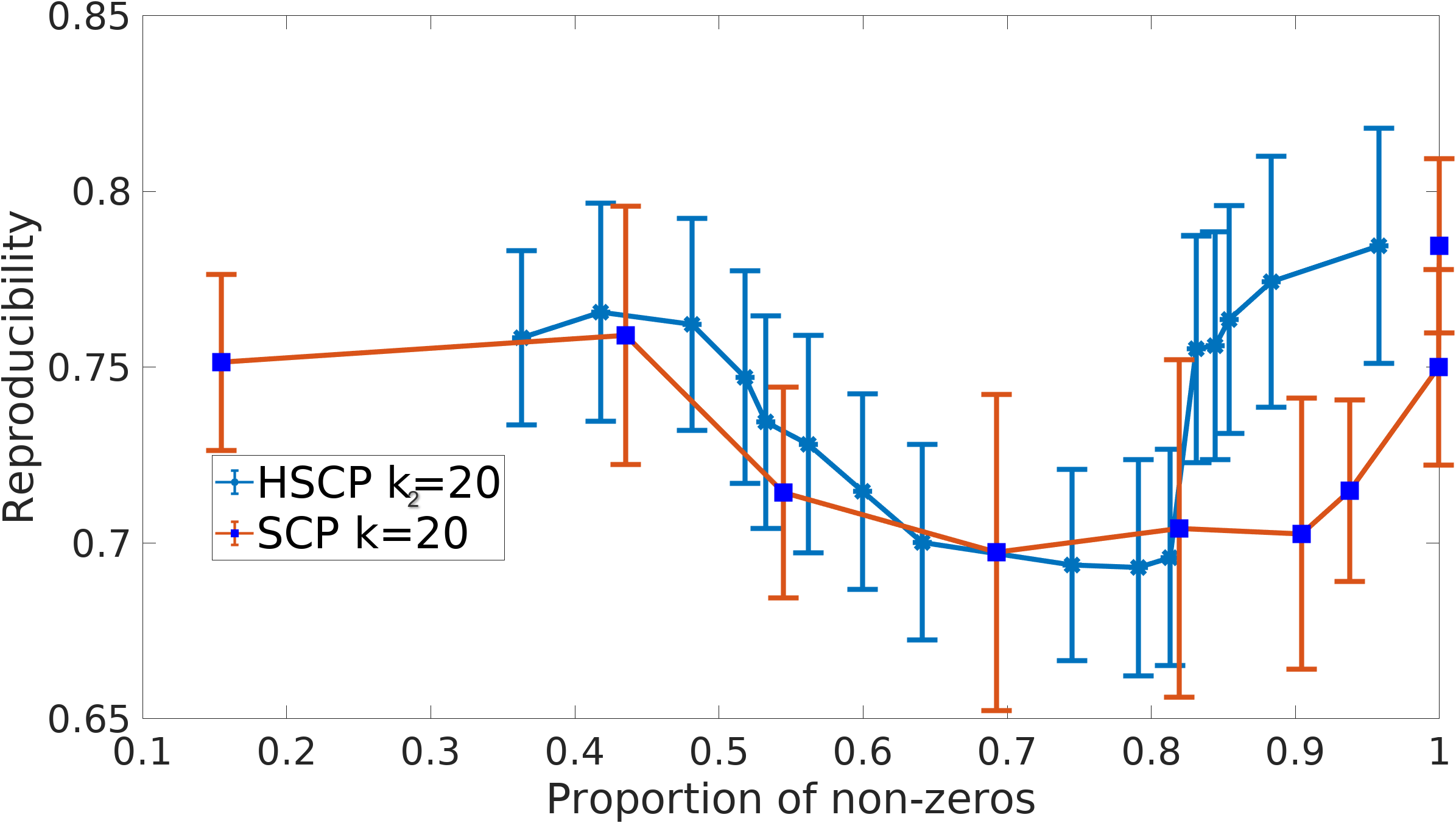}

    \end{minipage}
    \hfill
    \begin{minipage}[t]{0.49\textwidth}
        \centering
                \includegraphics[trim={0cm 0.1cm 0cm 0.1cm},clip,width=0.8\textwidth]{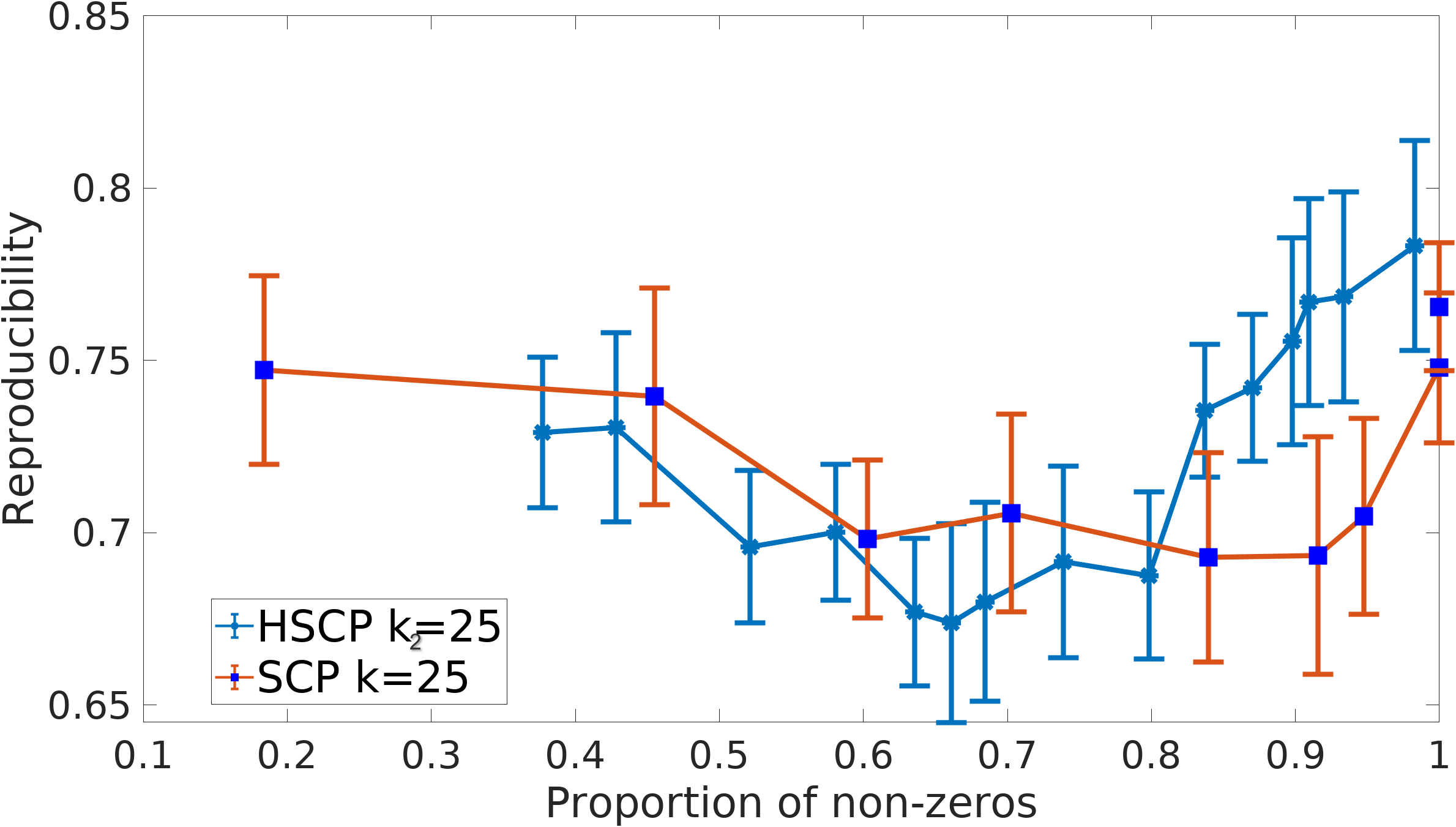}

    \end{minipage}

        \caption{Comparison of single scale (SCP) and hierarchical (HSCP) components on PNC dataset. X axis corresponds to proportion of non-zeros in the components. All the HSCP are second level components.}
    \label{fig:single_pnc}
    
\end{figure*}

\subsection{Comparison with single scale components}
\label{sssec:num1}
We also compared the reproducibility of the shared components extracted from the hierarchical model (hSCP) versus single scale components (SCP). Reproducibility here is defined as the normalized inner product of components derived from the two equal random sub-samples of the data averaged across all the components. We decomposed the correlation matrix into two levels to demonstrate the advantages of hierarchical factorization and show proof of concept. There might not be a single $K$ that best describes the data, and the algorithm allows us to investigate the continuum of functional connectivity patterns at different $K$s. We compare components derived from hSCP with $k_1 \in \{10,15,20,25\}$, $k_2\in\{4,5,6,8\}$, $\lambda_1 \in P\times 5(10^{[-3:-1]})$ and $\lambda_2 \in k_1\times10^{[-3:-1]}$, and from SCP with $k\in\{4,5,6,8,10,15,20,25\}$ at $\lambda \in P\times5(10^{[-4:-1]})$. At a fixed $k_2$ and $\lambda_2$, we find the optimal $k_1$ and $\lambda_1$ by dividing the data into three equal parts: training, validation, and test data, and choosing the parameters corresponding to maximum mean reproducibility over $20$ runs on training and validation set. Figure \ref{fig:single_hcp} and Figure \ref{fig:single_pnc} show the reproducibility of the components averaged over $20$ runs on training and test data. We can see that the same number of components extracted from the second level using hSCP are, on average, more reproducible than the components extracted using SCP.

\subsection{Comparison of hSCP with existing approaches}
We compared the reproducibility of hierarchical components extracted from hSCP to hierarchical overlapping communities obtained using EAGLE \cite{shen2009detect} and OSLOM \cite{lancichinetti2011finding}. Implementation of EAGLE and OSLOM was obtained from the authors. Correlation matrix averaged across all the subjects was used as an input to EAGLE and OSLOM. 
Hierarchical components and communities with $k_1 \in \{4,5,6,8\}$, $k_2\in\{10,15,20,25\}$ were generated from hSCP, EAGLE and OSLOM. Optimal $\lambda_1$ and $\lambda_2$ for hSCP were selected by dividing the data into three equal parts: training, validation and test set, and performing the validation procedure as described in section \ref{sssec:num1}. Reproducibility was computed using training and test for all the methods for all combinations of $k_1$ and $k_2$. Table\ref{table:hcp} and Table\ref{table:pnc} show the reproducibility results on HCP and PNC datasets. For a particular $k_1$ and $k_2$, reproducibility table show the average of the two reproducibility values. The results clearly show that the hSCPs have better reproducibility than the communities derived using EAGLE and OSLOM.

\begin{table}
  \begin{tabularx}{0.48\textwidth}{p{.07cm}p{0.7cm}p{2cm}p{2cm}p{2cm}}
    \toprule
    
    & & {$10$}   & {$15$} & {$20$} \\
    \midrule
    \multirow{ 3}{*}{$4$}& hSCP &     $0.8885\pm 0.0441$ &   $0.8351\pm 0.0748$ & $0.8507 \pm 0.0635$ 
    \\
    & EAGLE & $0.3077\pm 0.0981$  & $0.4158 \pm 0.1321$ & -      \\
    & OSLOM & $0.7493 \pm 0.0882$  & - & -      \\
    \midrule
        \multirow{ 3}{*}{$5$} & hSCP & $0.8753\pm 0.0348$ & $0.8356\pm 0.0591$ &   $0.8281 \pm 0.0656$   \\   
    & EAGLE & $0.2908 \pm 0.0737$  & $0.2664 \pm 0.0333$ & $0.0792\pm 0.1656$      \\
    & OSLOM & $0.6092\pm 0.0733$ & -& -      \\
    \midrule
        \multirow{ 3}{*}{$6$}& hSCP & $0.8756\pm 0.0375$ &  $0.8461 \pm 0.0486$ &   $0.8224 \pm 0.0555$ \\
   &EAGLE& $0.2356\pm 0.0196$  & $0.3209\pm 0.1206$ & $0.3717\pm 0.1698$       \\
   & OSLOM  & $0.5791\pm 0.0792$  & - & -     \\
   \midrule
       \multirow{ 3}{*}{$8$}& hSCP & $0.8781\pm 0.0694$ &    $0.8389\pm 0.0479$ &    $0.8240 \pm 0.0460$    \\
   &EAGLE & -  & - &  $0.3374 \pm 0.1672$      \\
    &OSLOM & -  & - &  -      \\
    
    \bottomrule
  \end{tabularx}\\
  \captionof{table}{{Reproducibility comparison (mean$\pm$std) on HCP dataset. The rows correspond to  values of $k_1$ and the columns correspond to values of $k_2$.}}
  \label{table:hcp}
\end{table}

\begin{table}
  \centering
 \begin{tabularx}{0.48\textwidth}{p{.07cm}p{0.7cm}p{2cm}p{2cm}p{2cm}}
    \toprule
    &     & {$10$}   & {$15$} & {$20$}  \\
    \midrule
    \multirow{ 3}{*}{$4$} & $hSCP$&    $0.8838\pm 0.0495$ &   $0.7998\pm 0.0766$  &  $0.8036\pm 0.0599$ 
    \\
    &EAGLE& $0.6287\pm 0.3005$  & $0.6433 \pm 0.1321$ & $0.6046\pm 0.2981$   \\
    &OSLOM& $0.6780\pm 0.0537$ & - & -   \\
    \midrule
        \multirow{ 3}{*}{$5$} &hSCP& $0.8785\pm 0.0675$  &  $0.8379\pm 0.0704$ &    $0.8099 \pm 0.0736 $      \\
    &EAGLE& $0.6575\pm 0.1973$  & $0.5327\pm 0.1828$ & $0.5426\pm 0.1656$       \\
    &OSLOM& $0.5867 \pm 0.0869$  & - & -       \\
    \midrule
        \multirow{ 3}{*}{$6$} &hSCP& $0.8655\pm 0.0404$  &  $0.8364\pm 0.0649$ &   $0.8518\pm 0.0587$  \\
     & EAGLE & $0.7571\pm 0.2366$  & $0.6279 \pm 0.1011$ & $0.6244 \pm 0.2627$      \\ 
     & OSLOM & $0.6391 \pm 0.1266$ & - & -      \\ 
     \midrule
       \multirow{ 3}{*}{$8$}&hSCP & $0.8670\pm 0.0559$ &    $0.8347\pm 0.0517$ &    $0.8340\pm 0.0657$   \\
      & EAGLE &-  & $0.7451 \pm 0.0319$ & $0.5933 \pm 0.2126$     \\
       & OSLOM & $0.5479 \pm 0.0987$ & - & -     \\
    
    \bottomrule
  \end{tabularx}
    \captionof{table}{{Reproducibility comparison (mean$\pm$std) on PNC dataset. The rows correspond to  values of $k_1$ and the columns correspond to values of $k_2$.}}
    \label{table:pnc}
\end{table}
\begin{table*}
  \centering
  \begin{tabular}{llllllllll}
    \toprule
    & &\multicolumn{4}{c}{Correlation} & \multicolumn{4}{c}{MAE (years)}\\
    &     & {$10$}   & {$15$} & {$20$} & {$25$}  & {$10$}   & {$15$} & {$20$} & {$25$} \\
    \midrule
    \multirow{ 3}{*}{$4$} & hSCP&    $0.259 \pm 0.010$ &   $0.301 \pm 0.014$  &  $0.319 \pm 0.012$ & $0.377 \pm 0.010$ & $3.20 \pm 0.06$ & $3.16 \pm 0.10$ & $3.13 \pm 0.09$ & $3.06 \pm 0.14$
    \\
    &EAGLE& $0.246 \pm 0.004$ &   $0.298 \pm 0.009$  &  $0.300 \pm 0.004$ & $0.347 \pm 0.005$ & $3.22 \pm 0.01$ & $3.20 \pm 0.01$ & $3.15 \pm 0.01$ & $3.10 \pm 0.01$   \\
     &OSLOM & $0.209 \pm 0.007$ &   -  &  - & -& $3.25 \pm 0.01$ & - & - & -   \\  
     \midrule
        \multirow{ 3}{*}{$5$} & hSCP&    $0.263 \pm 0.010$ &   $0.327 \pm 0.025$  &  $0.379 \pm 0.028$ & $0.403 \pm 0.017$ & $3.19 \pm 0.07$ & $3.10 \pm 0.17$ & $3.06 \pm 0.21$ & $3.03 \pm 0.13$
    \\
    &EAGLE& $0.259 \pm 0.003$ &   $0.298 \pm 0.002$ &  $0.301 \pm 0.006$ & - & $3.21 \pm 0.01$ & $3.13 \pm 0.01$ & $3.09 \pm 0.01$ & -   \\
      &OSLOM& $0.217 \pm 0.005$ &   -  &  - & - & $3.24 \pm 0.01$ & - & - & -   \\    
      \midrule
        \multirow{ 3}{*}{$6$} &hSCP& $0.257 \pm 0.013$  &  $0.342 \pm 0.027$ &   $0.381 \pm 0.021$ & $0.407 \pm 0.022$ & $3.20 \pm 0.08$ & $3.11 \pm 0.18$ & $3.06 \pm 0.16$  & $3.02 \pm 0.18$  \\
     & EAGLE & $0.281 \pm 0.004$  & $0.308 \pm 0.005$ & $0.321 \pm 0.007$ & - & $3.18 \pm 0.01$ & $3.15 \pm 0.01$  & $3.14 \pm 0.01$ & -\\ 
         &OSLOM& $0.236 \pm 0.008$ &   -  &  - & - & $3.20 \pm 0.01$ & - & - & -   \\
         \midrule
       \multirow{ 3}{*}{$8$}&hSCP & $0.278 \pm 0.022$ &    $0.372 \pm 0.026$ &  $0.382 \pm 0.023$ & $0.409 \pm 0.010$ & $3.18 \pm 0.14$ & $3.07 \pm 0.18$ & $3.05 \pm 0.17$  & $3.02 \pm 0.13$  \\
      & EAGLE & -  & $0.311 \pm 0.003$ & $0.326 \pm 0.007$  & - & - & $3.17 \pm 0.01$  & $3.13 \pm 0.02$  & - \\
          &OSLOM& $0.264 \pm 0.007$ &   -  &  - & - & $3.20 \pm 0.01$ & - & - & -   \\
    \bottomrule
  \end{tabular}
    \caption{Prediction performance comparison of hSCP, EAGLE and OSLOM}
    \label{table:pnc_age}
\end{table*}

\begin{table*}
  \centering
  \begin{tabular}{llllllllll}
    \toprule
    & &\multicolumn{4}{c}{Correlation} & \multicolumn{4}{c}{MAE (years)}\\
    &     & {$10$}   & {$15$} & {$20$} & {$25$}  & {$10$}   & {$15$} & {$20$} & {$25$} \\
    \midrule
    \multirow{ 2}{*}{$4$} &EAGLE& $0.0034$ &   $0.0229$ &  $3.6\times 10^{-4}$ & $4.7\times 10^{-5}$ & $0.0159$ & $0.0108$ & $0.0323$ & $0.0351$  \\
     &OSLOM & $4.7\times 10^{-5}$ &   -  &  - & -& $0.0012$ & - & - & -   \\  
     \midrule
        \multirow{ 2}{*}{$5$} &EAGLE& $0.0447$ &   $1.1\times 10^{-4}$  &  $4.7\times 10^{-5}$ & - & $0.0447$ & $0.1198$ & $0.0108$ & -   \\
      &OSLOM&  $4.7\times 10^{-5}$ &   -  &  - & - & $0.0413$ & - & - & -   \\    
      \midrule
        \multirow{ 2}{*}{$6$} & EAGLE & $1$  & $6.2\times 10^{-4}$ & $4.7\times 10^{-5}$ & - & $0.9727$ & $0.0653$  & $0.0145$ & -\\ 
         &OSLOM& $1.3\times 10^{-4}$ &   -  &  - & - & $0.778$ & - & - & -   \\
         \midrule
       \multirow{ 2}{*}{$8$}& EAGLE & -  & $4.7\times 10^{-5}$ & $4.7\times 10^{-5}$  & - & - & $0.0447$  & $0.0209$ & - \\
          &OSLOM& $0.0175$ &   -  &  - & - & $0.0563$ & - & - & -   \\
    \bottomrule
  \end{tabular}
    \caption{{p-value from Wilcoxon signed-rank test on different performance measures}}
    \label{table:pnc_age_stats}
\end{table*}
\subsection{Age prediction}
We compared the predictability power on the age prediction problem of the hierarchical components extracted from hSCP, EAGLE and OSLOM. Using PNC dataset, we first extracted the components and their strength ($\mathbf{\Lambda}$) for each individual. These strength values were then used to predict age of each individual using linear regression. Pearson correlation coefficient and mean absolute error (MAE) between the predicted brain age and the true age was used as the performance measure for comparison. Table \ref{table:pnc_age} summarizes the result obtained. {To determine if our results are significantly better, the Wilcoxon signed-rank test was performed as the information about the underlying distribution in case of different performance measures is unknown. As the lower MAE is preferred, we performed a left-tailed hypothesis test when MAE is used as a performance measure. A right-tailed hypothesis test is performed when correlation is used as a performance measure because a higher value of correlation is better. Below is the null hypotheses in the two case:
\begin{itemize}
    \item There is no difference between correlation values obtained from our method compared to other methods
    \item There is no difference between MAE values obtained from our method compared to other methods
\end{itemize}
Table \ref{table:pnc_age_stats} demonstrates that the prediction model built using hSCPs performed significantly better (p-value $< 0.05$) better than the model built using EAGLE and OSLOM components in the majority of the cases. This indicates that the hSCPs were more informative for predicting brain age. One of the reasons for the poor performance of EAGLE was that it only estimated if a region is present or not present in a component. In contrast, hSCP can determine the strength of the presence, thus had more degree of freedom resulting in better performance.}

\subsection{Clustering}
An extension of the above method is presented below, which estimates hSCPs for better clustering of the data and capturing of heterogeneity. Data clustering is performed using subject specific information of the components. We add a penalty term for clustering in the objective function given in problem \ref{problem1}. The modified objective function is given in problem \ref{problem2}. The joint minimization problem for estimating hSCPs and using their subject specific information for clustering is given below: 
 \begin{equation}
\begin{aligned}
& \underset{\mathcal{W},\mathcal{L},\mathcal{C}}{\text{minimize}}
& & H(\mathcal{W},\mathcal{L}) + \sum_{r=1}^{K}\sum_{l=1}^{L}\sum_{i \in \mathcal{M}_l}\sum_{d=1}^{k_r}||\frac{{x}^i_{d,r}}{\|\mathbf{x}_{d,r} \|} - {c}_{d,r}^l ||^2\\
& \text{subject to}
&&\|\mathbf{w}^r_l\|_1 < \lambda_r, l=1,...,k_r \hspace{0.3cm} \text{and} \hspace{0.3cm} r=1,..,K  \\
&&&\|\mathbf{w}^r_l\|_\infty \leq 1, l=1,...,k_r \hspace{0.3cm} \text{and} \hspace{0.3cm} r=1,..,K \\
&&&\mathbf{W}_j \geq 0, j=2,...,K\\
&&& \mathbf{\Lambda}_r^i \succeq 0, i=1,...,S  \hspace{0.3cm} \text{and} \hspace{0.3cm} r=1,..,K \\
&&& \trace(\mathbf{\Lambda}_r^i) =1, i=1,...,S  \hspace{0.3cm} \text{and} \hspace{0.3cm} r=1,..,K 
\end{aligned}
   \label{problem2}
 \end{equation}
  \begin{algorithm}[t]
 \caption{hSCP-clust}\label{alg:HSCP-clust}
\begin{algorithmic}[1]
\State \textbf{Input:} Data $\Theta$, number of connectivity patterns $k_1$,..,$k_K$ and sparsity $\lambda_1$,..,$\lambda_K$ at different level 
\State $\mathcal{W}$ and $\mathcal{L}$ = Initialization($\Theta$) 
\State random initialization for $\mathcal{}$ (uniform sampling in [0, 1])
\Repeat 
\For{$r=1$ {\bfseries to} $K$} 
\State Step 5-9 from Alg \ref{alg:HSCP}
\For{$i= 1,..,S$} 
\State $\mathbf{\Lambda}_r^i \leftarrow \descent(\mathbf{\Lambda}_r^i)$
\State $\mathbf{\Lambda}_r^i \leftarrow \proj_2(\mathbf{\Lambda}_r^i) $
\EndFor
\State $\mathcal{C}, \{\mathcal{M}_l\}_{l=1:L} \leftarrow $  k-means($\mathbf{\Lambda}_r^i$)
\EndFor
\Until{Stopping criterion is reached}
\State \textbf{Output:} $\mathcal{W}$, $\mathcal{L}$, $\mathcal{C}$ and $\{\mathcal{M}_l\}_{l=1:L}$
\end{algorithmic}
\end{algorithm}
    \begin{figure}
    \begin{minipage}[t]{.47\textwidth}
        \centering
        \includegraphics[width=\textwidth]{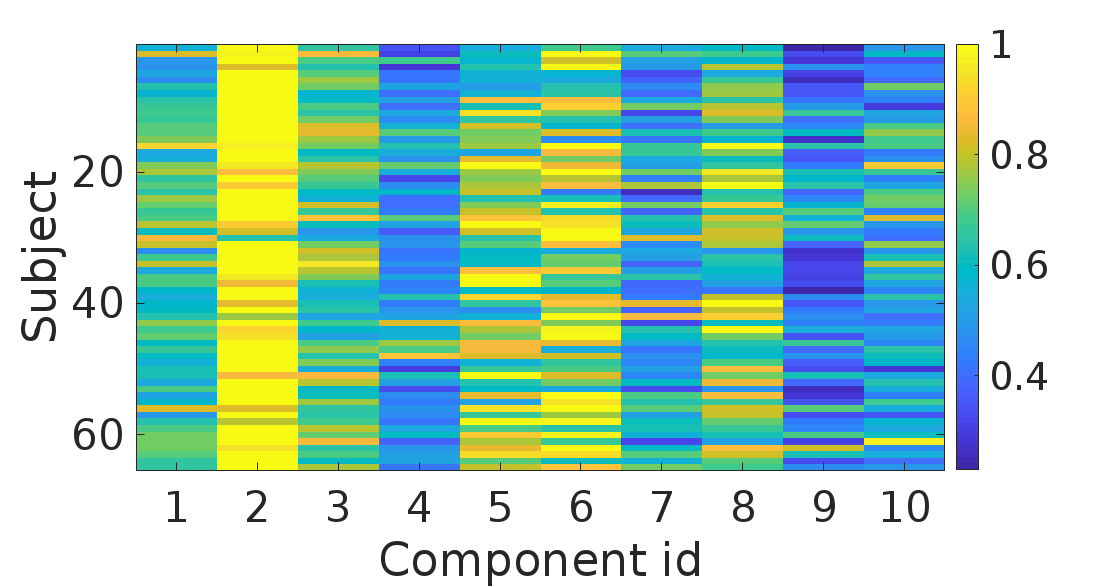}
        \subcaption{Heterogeneity captured by subjects of cluster 1}
    \end{minipage}
    \begin{minipage}[t]{.47\textwidth}
        \centering
        \includegraphics[width=\textwidth]{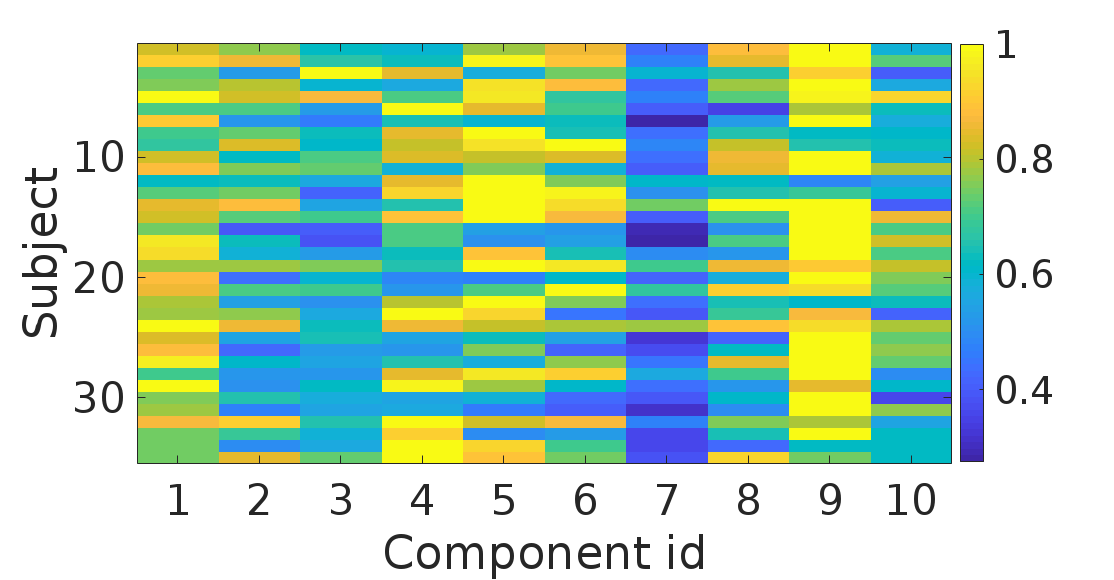}
        \subcaption{Heterogeneity captured by subjects of cluster 2}
    \end{minipage}

        \caption{Heterogeneity captured by fine scale components in HCP. The color indicates the strength ($\mathbf{\Lambda}_2$) of each component present in a subject. Maximum strength of a component across subjects  is fixed to be 1 for comparison purpose.}
    \label{fig:hetero}
\end{figure}
    \begin{figure*}
    \begin{minipage}[t]{.47\textwidth}
        \centering
        \includegraphics[width=\textwidth]{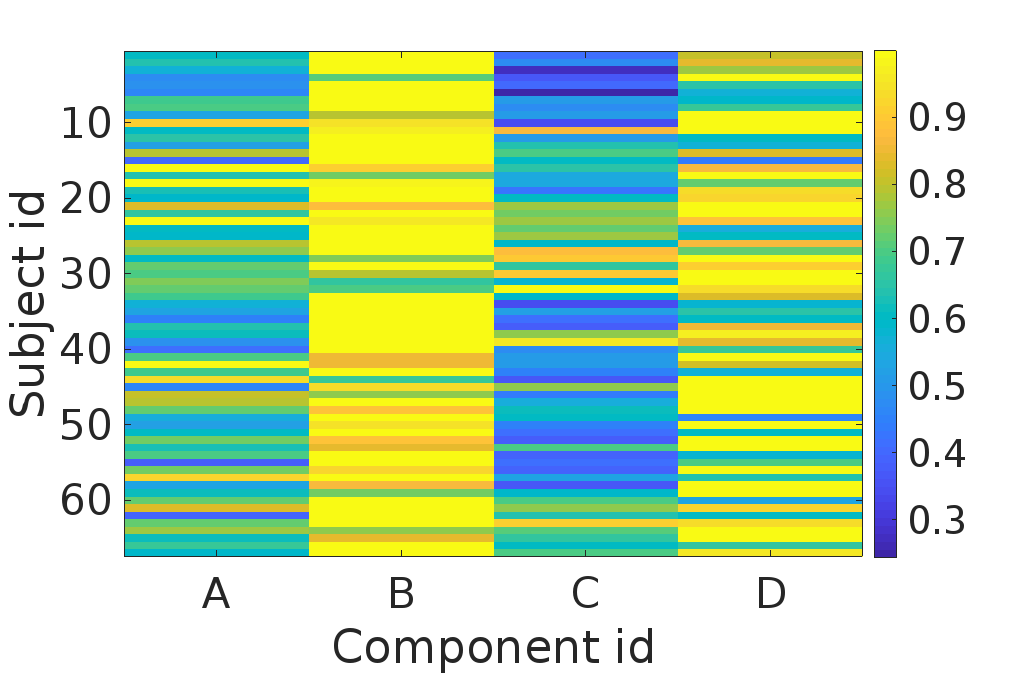}
        \subcaption{Heterogeneity captured by subjects of cluster 1}
    \end{minipage}
    \begin{minipage}[t]{.47\textwidth}
        \centering
        \includegraphics[width=\textwidth]{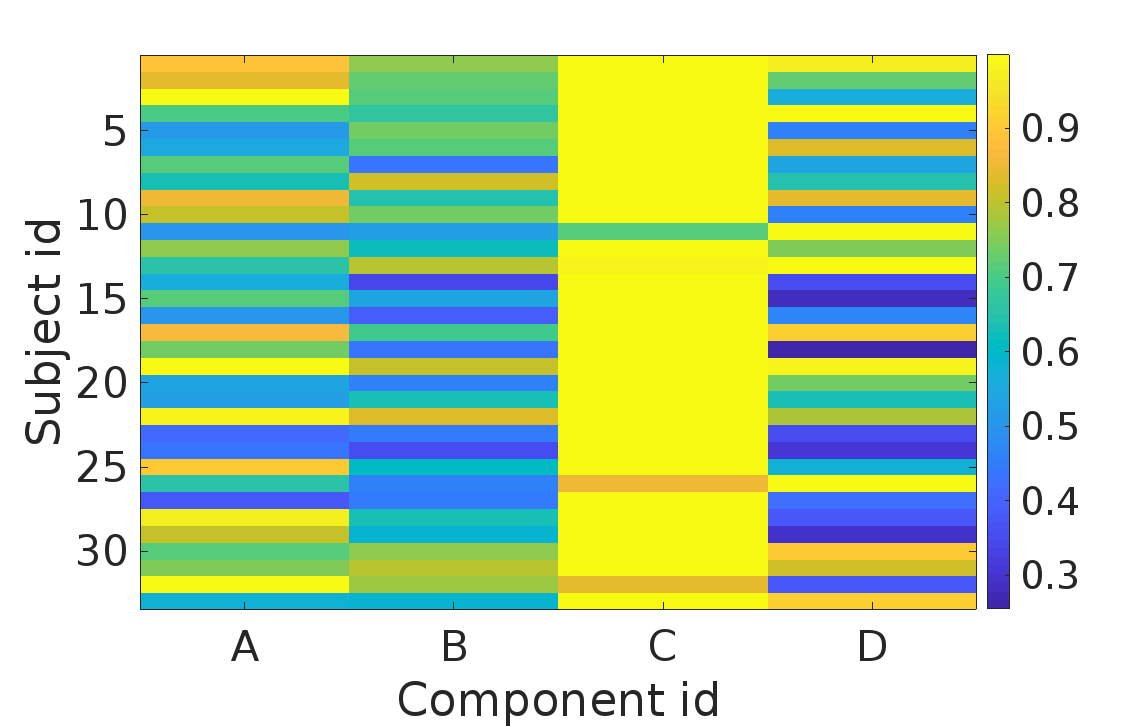}
        \subcaption{Heterogeneity captured by subjects of cluster 2}
    \end{minipage}

        \caption{Heterogeneity captured by coarse scale components in HCP. The color indicates the strength ($\mathbf{\Lambda}_1$) of each component present in a subject. Maximum strength of a component across subjects  is fixed to be 1 for comparison purpose.}
    \label{fig:hetero1}
\end{figure*}
 where $\mathbf{c}^l_r$ is the $l^{th}$ cluster center at the $r^{th}$ level, $L$ is the number of clusters, $\mathcal{C} = \{\mathbf{c}^l_r\}_{l=1:L, r = 1:K}$, $\mathbf{x}^i_r$ stores the diagonal elements of $\mathbf{\Lambda}^i_r$ and $\mathcal{M}_l$ stores the information whether $i^{th}$ subject belongs to $l^{th}$ cluster or not. In the above problem, $||\frac{{x}^i_{d,r}}{\|\mathbf{x}_{d,r} \|} - {c}_{d,r}^l ||^2$ penalty is used for incorporating clustering by penalizing distance between points in a cluster and cluster center, and ${x}^i_{d,r}$ is divided by $\|\mathbf{x}_{d,r} \|$ for normalization. The above non-convex problem can be solved in a similar way as the problem \ref{problem1} is solved in section \ref{hSCP} using alternating minimization. Alg \ref{alg:HSCP-clust} provides a complete procedure for solving the problem. In the Alg \ref{alg:HSCP-clust}, k-means($\mathbf{\Lambda}_r^i$) is used for applying k-means clustering \cite{wagstaff2001constrained} on $\mathbf{\Lambda}_r^i$ and k-means($\mathbf{\Lambda}_r^i$) outputs $\mathcal{C}$ as cluster centers and $\{\mathcal{M}_l\}_{l=1:L}$ as cluster assignments of $\mathcal{L}$. We ran the algorithm on HCP data to extract the components and the clusters in the data. Number of clusters was selected by first extracting the hierarchical components without the penalty term and then clustering the data by using k-means on $\mathcal{L}$. $L$ which is number of clusters was set to 2 by using the elbow method. Number of of coarse scale components was set to be 4 and and fine scale components to be 10 since they exhibited the highest reproducibility between the training and test sets. Figure \ref{fig:hetero} and Figure \ref{fig:hetero1} show the distribution of fine and coarse components in two clusters. From Figure \ref{fig:hetero1}, we can see that component A  and C are more prominent in cluster 2 compared to cluster 1, and component B and D are prominent in cluster 1 compared to cluster 2. The algorithm has forced Component A and its sub-components to have higher weights in one cluster. But for component B, sub-components 2 and 3 are prominent in cluster 1 and sub-component 1 is prominent in cluster 2 which can be seen in Figure \ref{fig:hetero}. From the Figure \ref{fig:hetero} and \ref{fig:hetero1}, it can be seen that our method can reveal heterogeneity in the population by capturing the strength of components' presence in each individual.

\subsection{Results from resting state fMRI}
Figure \ref{fig:comp} displays the 10 fine level components, the 4 coarse level components and the hierarchical structure. It can be clearly seen from Figure \ref{fig:comp} that the fine and coarse level components are overlapping and sparse, and coarse components are comprised of a sparse linear combination of fine level components which helps in discovering the relation between different networks at different scales. 

{The ten fine level components obtained show the relation between different functional networks and are similar to the SCPs extracted in \cite{eavani2015identifying}. From Fig. \ref{fig:comp}, it can be seen that our approach can separate task-positive regions and their associated task-negative regions into separate patterns without using traditional seed-based methods that require knowledge of a seed region of interest. Various studies have found that task-positive regions are positively correlated with each other, and task-negative regions are positively correlated with each other. The regions in the two networks are negatively correlated with each other, which aligns with our results \cite{raichle2015brain,yeo2014estimates}. Component 2 covers Default Mode Network and Dorsal Attention Network, which are anti-correlated with each other. This result is a well-known finding, previously described using the seed-based correlation method \cite{fox2005human}. Anti correlations between different brain regions can represent interactions that are dependent on the state of the brain. As our method is not capturing dynamics, it has captured the interactions between different regions in different components. An example of this anti-correlation between the default mode network and the task positive network; these interactions are thought to be facilitated by indirect anatomical connections between the regions of two networks\cite{buckner2013opportunities}. Component 8 shows different regions of DMN anti-correlated with sensorimotor, described in separate study \cite{karahanouglu2015transient}. Component C comprises of three connectivity patterns that involve the sensorimotor areas and its anti-correlations. Component A consists of Visual Network and Ventral Attention Network, which are anti-correlated with each other. 
From Fig. \ref{fig:comp}, we can see that part of sensorimotor and emotion networks \cite{drevets1998reciprocal} are anti-correlated with each other. These connections highlighted by our method are corroborated by the fact that these regions have direct anatomical pathways \cite{vergani2014white}. These negatively correlated networks can highlight different interactions in different brain regions, such as suppression, inhibition, and neurofeedback.  An extension of this method that estimates dynamic components can help us understand different anti-correlations mechanism between the regions. Future research is needed to understand more about the anti-correlation and the source of these interactions.
\begin{figure*}[h!tbp]
\begin{subfigure}{0.48\textwidth}
    \centering
\begin{minipage}[ct]{\textwidth}
\centering
\begin{forest}
  styleA/.style={top color=white, bottom color=white},
  styleB/.style={%
    top color=white,
    bottom color=red!20,
    delay={%
      content/.wrap value={##1\\{\includegraphics[scale=.5]{hand}}}
    }
  },
  for tree={
    rounded corners,
    draw,
    align=center,
    top color=white,
    bottom color=blue!20,
  },
  forked edges,
  [A{ \includegraphics[trim={1cm 3cm 1cm 2cm},clip,scale=0.2]{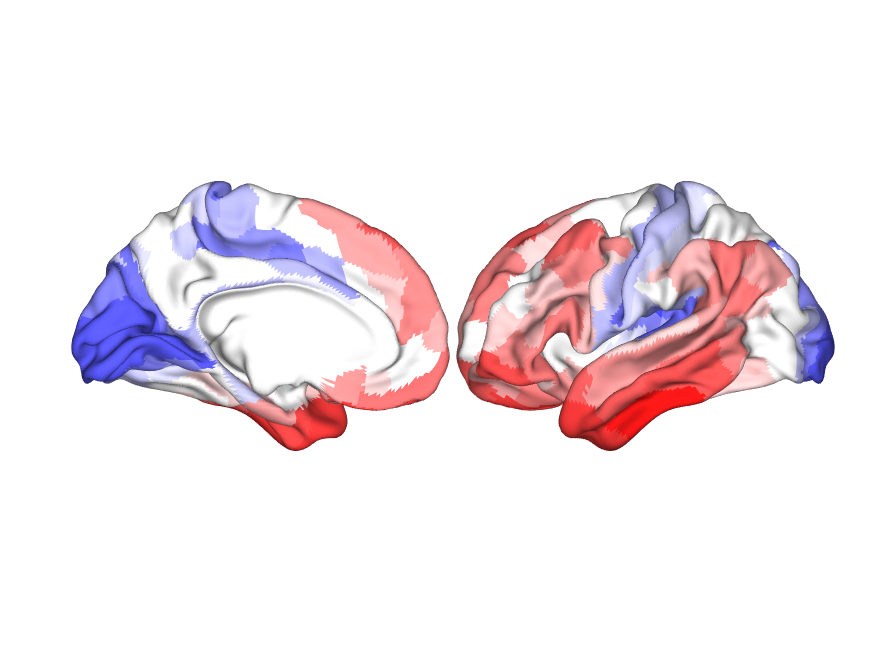}},styleA
    [1{ \includegraphics[trim={4.5cm 1.1cm 4cm 0.8cm},clip,scale=0.24]{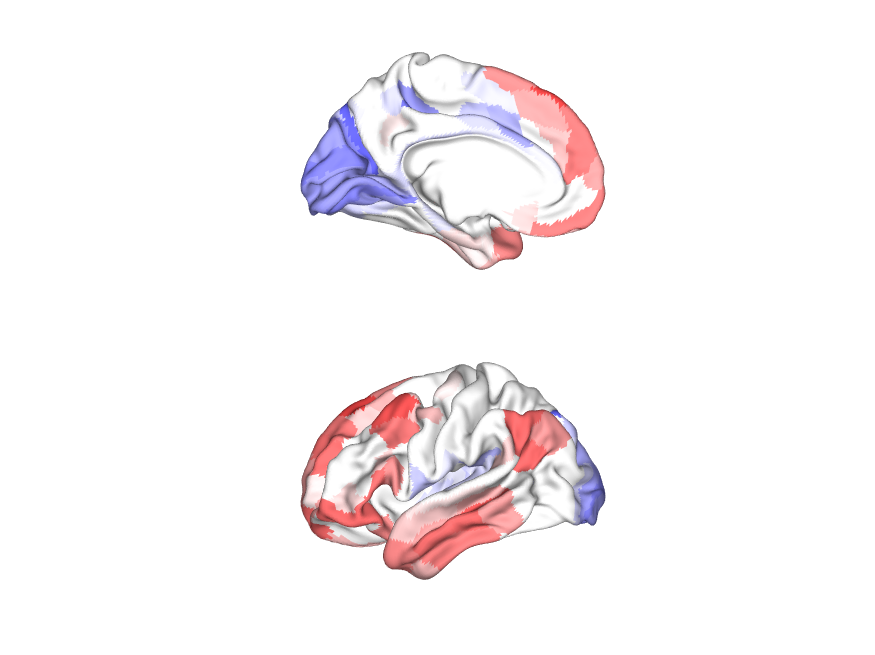}},styleA]
    [9{ \includegraphics[trim={4.5cm 1.1cm 4cm 0.8cm},clip,scale=0.24]{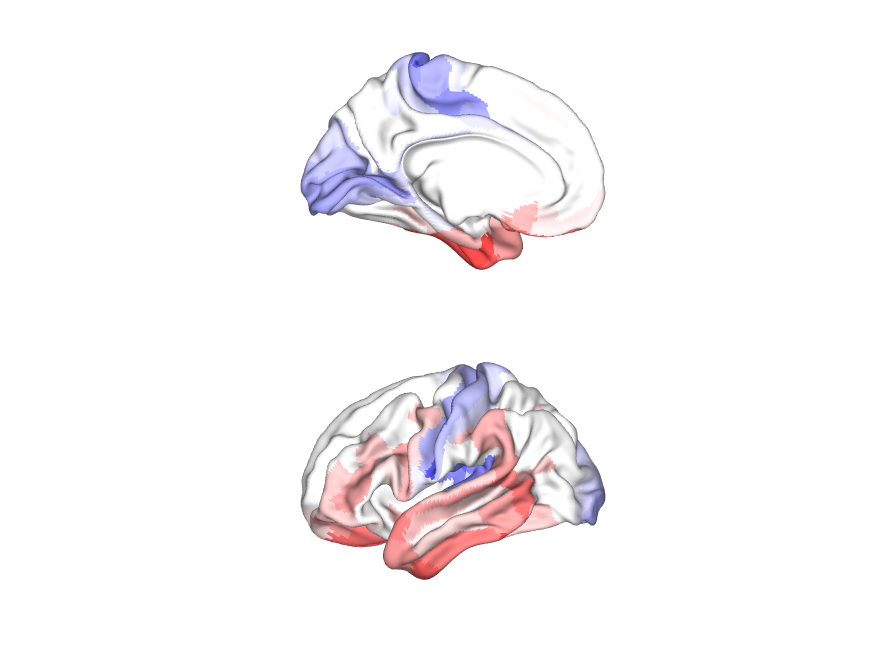}},styleA]
     ]
  ]
\end{forest}
\caption{Component A comprises of 1 and 9}
\end{minipage}
\end{subfigure}
\vspace{1em}
\hfill
\begin{subfigure}{0.48\textwidth}
\begin{minipage}[ct]{\textwidth}
\centering
\begin{forest}
  styleA/.style={top color=white, bottom color=white},
  styleB/.style={%
    top color=white,
    bottom color=red!20,
    delay={%
      content/.wrap value={##1\\{\includegraphics[scale=.5]{hand}}}
    }
  },
  for tree={
    rounded corners,
    draw,
    align=center,
    top color=white,
    bottom color=blue!20,
  },
  forked edges,
  [B{ \includegraphics[trim={1cm 3cm 1cm 2cm},clip,scale=0.22]{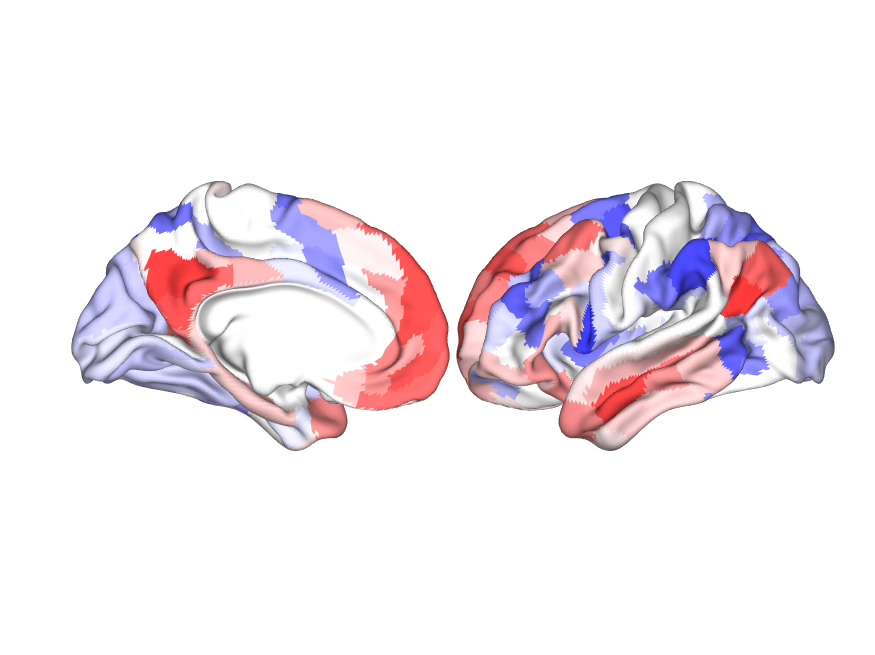}},styleA
    [1{ \includegraphics[trim={4.5cm 1.1cm 4cm 0.8cm},clip,scale=0.24]{figures/brain4/sahoo17.png}},styleA]
    [2{ \includegraphics[trim={4.5cm 1.1cm 4cm 0.8cm},clip,scale=0.24]{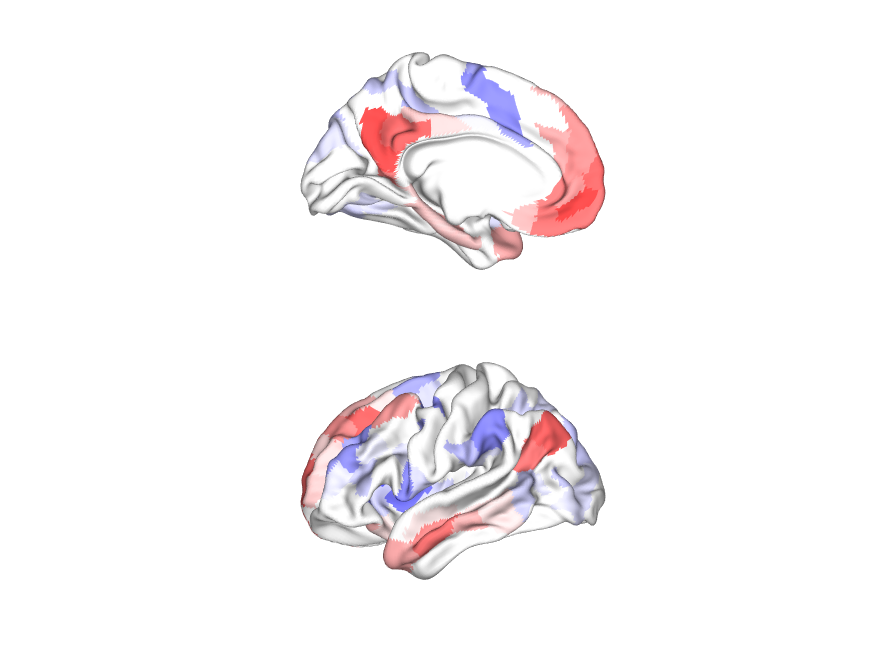}},styleA]
     [3{ \includegraphics[trim={4.5cm 1.1cm 4cm 0.8cm},clip,scale=0.24]{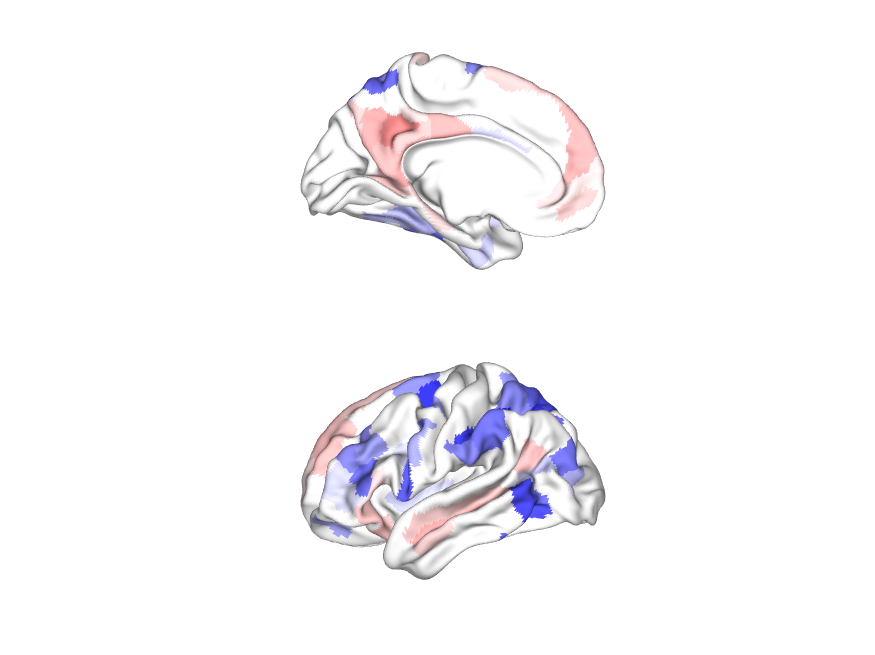}},styleA
        ]
     ]
  ]
\end{forest}
\caption{Component B comprises of 1, 2 and 3}
\end{minipage}
\end{subfigure}
\vspace{1em}
\begin{subfigure}{0.52\textwidth}
    \centering
\begin{minipage}[ct]{\textwidth}
\begin{forest}
  styleA/.style={top color=white, bottom color=white},
  styleB/.style={%
    top color=white,
    bottom color=red!20,
    delay={%
      content/.wrap value={##1\\{\includegraphics[scale=.5]{hand}}}
    }
  },
  for tree={
    rounded corners,
    draw,
    align=center,
    top color=white,
    bottom color=blue!20,
  },
  forked edges,
  [C{ \includegraphics[trim={1cm 3cm 1cm 2cm},clip,scale=0.22]{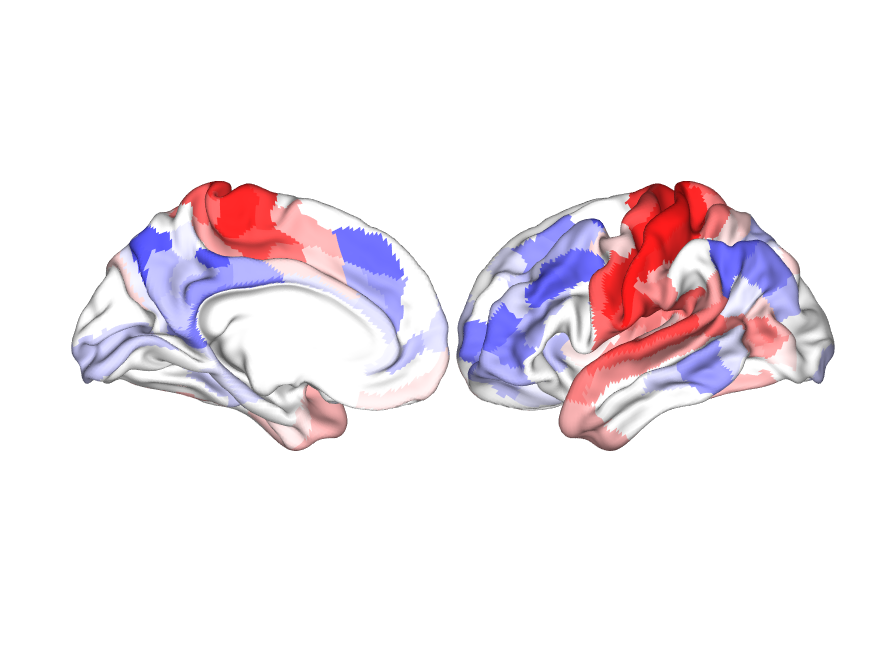}},styleA
    [6{ \includegraphics[trim={4.5cm 1.1cm 4cm 0.8cm},clip,scale=0.24]{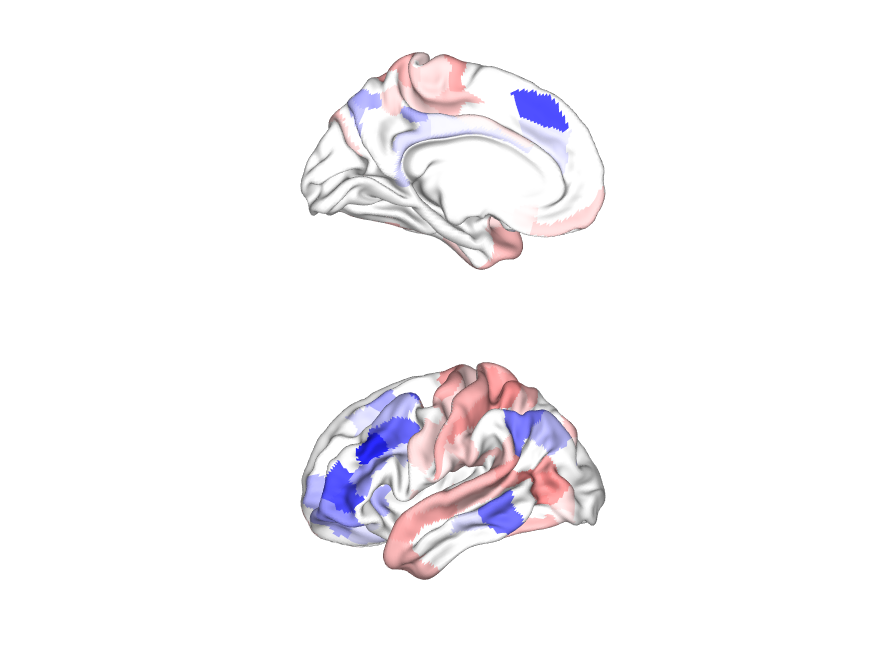}},styleA]
    [7{ \includegraphics[trim={4.5cm 1.1cm 4cm 0.8cm},clip,scale=0.24]{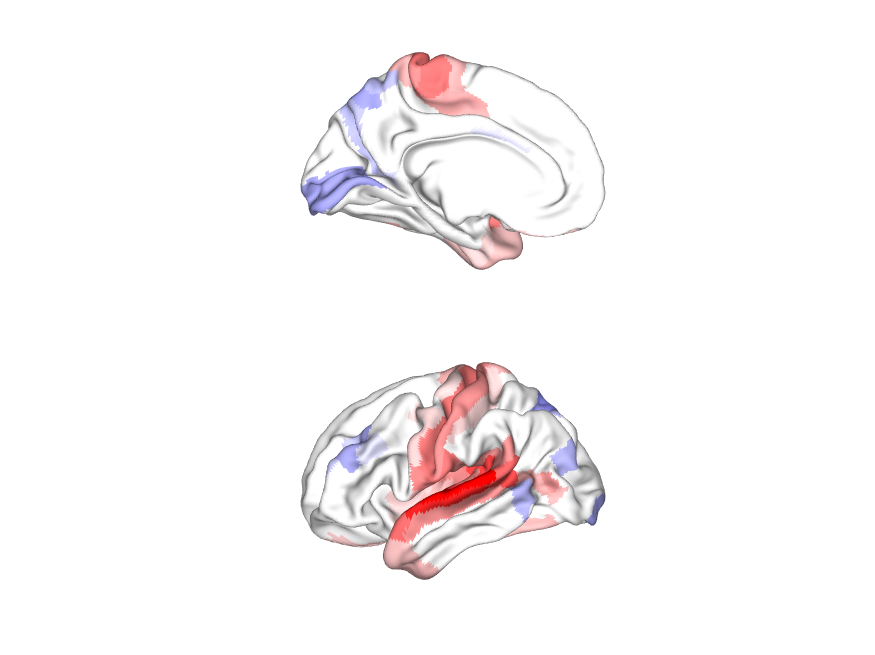}},styleA]
     [8{ \includegraphics[trim={4.5cm 1.1cm 4cm 0.8cm},clip,scale=0.24]{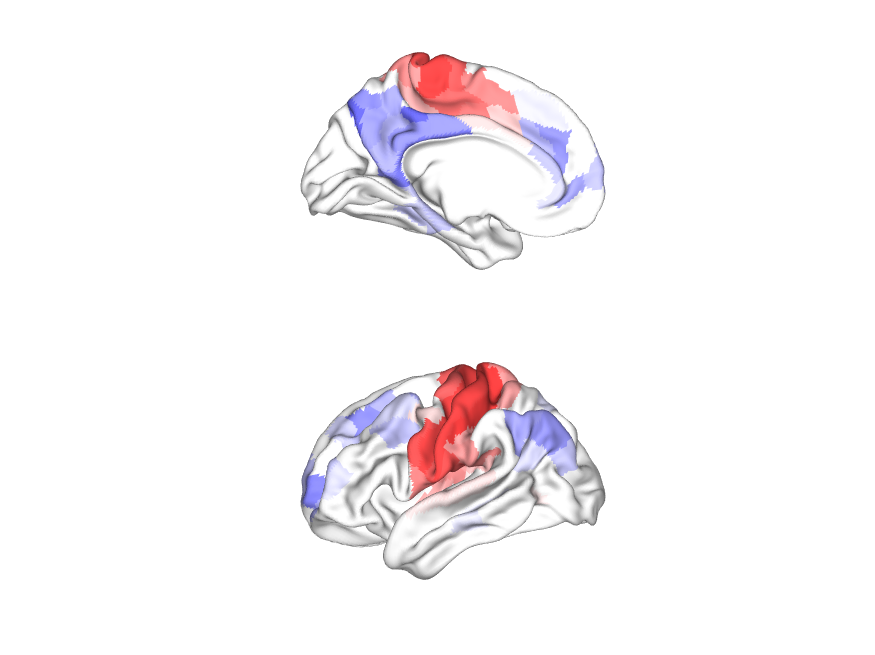}},styleA]
    [10{ \includegraphics[trim={9.4cm 2cm 7.9cm 1.2cm},clip,scale=0.11]{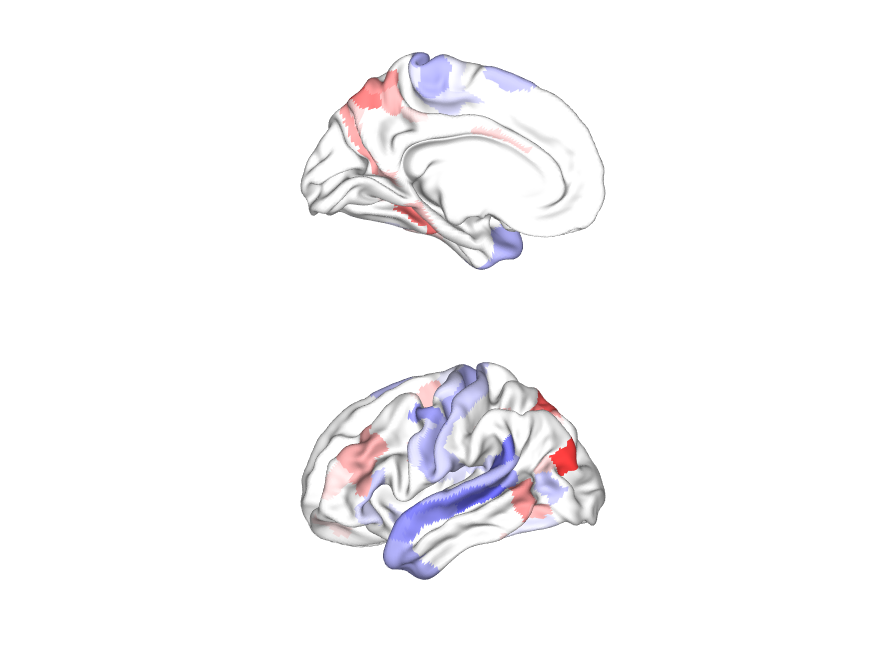}},styleA
        ]
     ]
  ]
\end{forest}
\caption{Component C comprises of 6, 7 and 8}
\end{minipage}
\end{subfigure}
\begin{subfigure}{0.48\textwidth}
    \centering
\begin{minipage}[ct]{\textwidth}
\centering
\begin{forest}
  styleA/.style={top color=white, bottom color=white},
  styleB/.style={%
    top color=white,
    bottom color=red!20,
    delay={%
      content/.wrap value={##1\\{\includegraphics[scale=.5]{hand}}}
    }
  },
  for tree={
    rounded corners,
    draw,
    align=center,
    top color=white,
    bottom color=blue!20,
  },
  forked edges,
[D{ \includegraphics[trim={1cm 3cm 1cm 2cm},clip,scale=0.22]{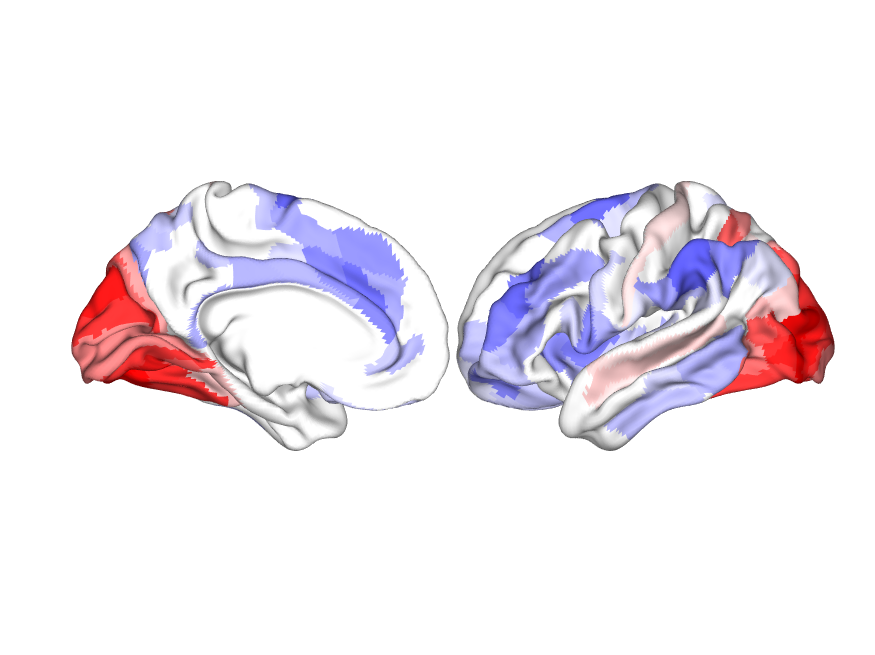}},styleA
    [4{ \includegraphics[trim={4.5cm 1.1cm 4cm 0.8cm},clip,scale=0.24]{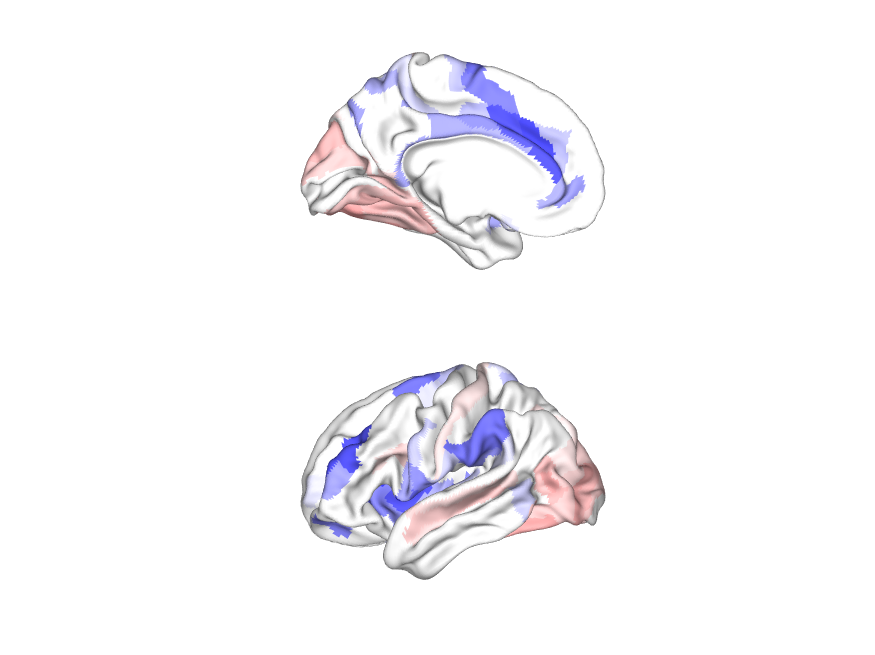}},styleA]
    [5{ \includegraphics[trim={4.5cm 1.1cm 4cm 0.8cm},clip,scale=0.24]{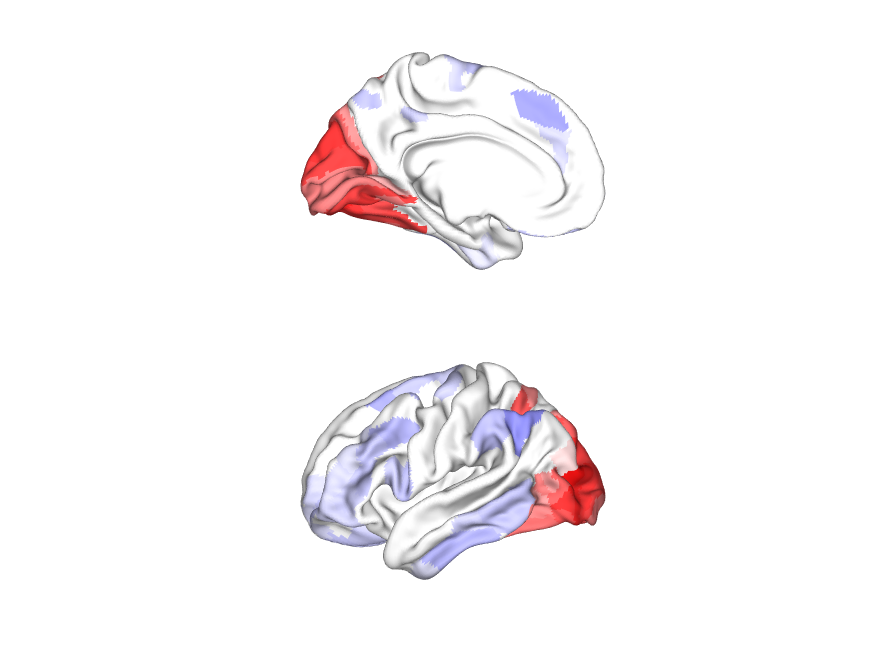}},styleA]
     [6{ \includegraphics[trim={4.5cm 1.1cm 4cm 0.8cm},clip,scale=0.24]{figures/brain4/sahoo22.png}},styleA
        ]
     ]
  ]
\end{forest}
\caption{Component D comprises of 4, 5 and 6}
\end{minipage}
\end{subfigure}
   \caption{Hierarchical components derived from HCP dataset showing the connection between 10 fine scale components ($\mathbf{W}_1$) denoted from 1 to 10 and 4 coarse scale components ($\mathbf{W}_1\mathbf{W}_2$) denoted from A to D. Nodes with red and blue color are correlated among themselves, but are anitcorrelated with each other. Note that blue color does not need to be necessarily associate with positive or negative correlation because the colors can be flipped without affecting the solution. 8 fine scale components ($\mathbf{W}_1$). 2 and 3 show different regions of Default Mode anti-correlated with the y Dorsal Attention and Cingulo-Opercular system. 8 show different regions of default mode anti-correlated with the sensori-motor areas. 4 and 5 shows different regions of Visual system anti-correlated with Salience and fronto-parietal control systems.}
    \label{fig:comp}
\end{figure*}

Our study also finds that compared to the primary sensory cortex, the higher-order association cortex has more has more associations in different components, shown in previous studies \cite{geranmayeh2014overlapping,beldzik2013contributive}. Traditional seed-based approaches have been used to show that these regions have functional connectivity with more heterogeneous regions implying that they receive input and send outputs to more diverse brain regions \cite{katsuki2012unique,crossley2014hubs}. Thus, allowing overlapping components and positive and negative correlations within the same components provides additional insights. These features of the method facilitate storing the relation of various overlapping regions within a functional system with other areas by assigning them to different components.}

Another important observation is that each of the coarse components comprises fine level components having major functional networks and their relation with other nodes. For instance, coarse component B includes majorly of 2 and 3, which stores the link between regions of Default Mode network and other nodes in the brain. Similarly, coarse component D saves the relationship between visual areas and the rest of the brain regions using 4 and 5. Thus, hSCPs can provide novel insights into the functioning of the brain by jointly uncovering both fine and coarse level components with the coarse components comprised of similarly functioning fine components.

\section{DISCUSSION AND CONCLUSIONS}
In this work, we proposed a novel technique for hierarchical extraction of  sparse components from correlation matrices, with application to rsfMRI data . The proposed method is a cascaded joint matrix factorization problem where correlation matrix corresponding to each individual's data is considered as an independent observation, thus allowing us to model inter-subject variability. We formulated the problem as non-convex optimization with sparsity constraints. It is important to note  that as the decomposition is not by itself unique, the ability to reproducibly recover components hinges on imposing sparsity in the decomposition which appears to provide useful and reproducible representations. We used adaptive learning rate based gradient descent algorithm to solve the optimization problem. Compared to the implementation of SCP, which had random initialization, we used SVD initialization which made the complete algorithm both deterministic and faster.

{In addition to shared patterns, we are able to extract the `strength' of these patterns in individual components, thus capturing heterogeneity across data.} Experimentally we showed that our method is able to find sparse, low rank hierarchical decomposition using cascaded matrix factorization which is highly reproducible across data-sets. Experimental results using the PNC dataset demonstrate that the hierarchical components extracted using our model could better predict the brain age compared to EAGLE and OSLOM. We also show that our model is able to capture heterogeneity using HCP dataset. Our model computationally extracts a set of hierarchical components common across subjects, including resting state networks. At the same time, we capture individual information about subjects as a linear combination of these hierarchical components, making it a useful measure for group studies. Importantly, our work provides a method to uncover hierarchical organization in the functioning of the human brain.  

There are several directions for the future work. Firstly, it is possible to extend the idea to estimate dynamic hierarchical components similar to \cite{cai2017estimation} which can help reveal how the hierarchical networks are varying over time. Secondly, generative-discriminative models can be build on the top of hSCP to find the components which are highly discriminative of some particular groups. For example, such model can estimate the hierarchical components which are most discirimanative of a neurodegenerative disorder. Third, it would be interesting to find the guarantee on the estimation error of the hierarchical components. One possible approach is to adapt the proof techniques of \cite{yu2018recovery}. Finally, future studies incorporating cognitive, clinical, and genetic data, might elucidate the biological underpinning and clinical significance of the heterogeneity captured by our approach. Such studies are beyond the scope of the current paper.  

\bibliographystyle{IEEEtran}
\bibliography{example_paper.bib}

\begin{thebibliography}{10}
\providecommand{\url}[1]{#1}
\csname url@samestyle\endcsname
\providecommand{\newblock}{\relax}
\providecommand{\bibinfo}[2]{#2}
\providecommand{\BIBentrySTDinterwordspacing}{\spaceskip=0pt\relax}
\providecommand{\BIBentryALTinterwordstretchfactor}{4}
\providecommand{\BIBentryALTinterwordspacing}{\spaceskip=\fontdimen2\font plus
\BIBentryALTinterwordstretchfactor\fontdimen3\font minus
  \fontdimen4\font\relax}
\providecommand{\BIBforeignlanguage}[2]{{%
\expandafter\ifx\csname l@#1\endcsname\relax
\typeout{** WARNING: IEEEtran.bst: No hyphenation pattern has been}%
\typeout{** loaded for the language `#1'. Using the pattern for}%
\typeout{** the default language instead.}%
\else
\language=\csname l@#1\endcsname
\fi
#2}}
\providecommand{\BIBdecl}{\relax}
\BIBdecl

\bibitem{sporns2010networks}
O.~Sporns, \emph{Networks of the Brain}.\hskip 1em plus 0.5em minus 0.4em\relax
  MIT press, 2010.

\bibitem{doucet2011brain}
G.~Doucet, M.~Naveau, L.~Petit, N.~Delcroix, L.~Zago, F.~Crivello, G.~Jobard,
  N.~Tzourio-Mazoyer, B.~Mazoyer, E.~Mellet \emph{et~al.}, ``Brain activity at
  rest: a multiscale hierarchical functional organization,'' \emph{Journal of
  neurophysiology}, vol. 105, no.~6, pp. 2753--2763, 2011.

\bibitem{park2013structural}
H.-J. Park and K.~Friston, ``Structural and functional brain networks: from
  connections to cognition,'' \emph{Science}, vol. 342, no. 6158, p. 1238411,
  2013.

\bibitem{ferrarini2009hierarchical}
L.~Ferrarini, I.~M. Veer, E.~Baerends, M.-J. van Tol, R.~J. Renken, N.~J.
  van~der Wee, D.~J. Veltman, A.~Aleman, F.~G. Zitman, B.~W. Penninx
  \emph{et~al.}, ``Hierarchical functional modularity in the resting-state
  human brain,'' \emph{Human brain mapping}, vol.~30, no.~7, pp. 2220--2231,
  2009.

\bibitem{meunier2009hierarchical}
D.~Meunier, R.~Lambiotte, A.~Fornito, K.~Ersche, and E.~T. Bullmore,
  ``Hierarchical modularity in human brain functional networks,''
  \emph{Frontiers in neuroinformatics}, vol.~3, p.~37, 2009.

\bibitem{beckmann2005investigations}
C.~F. Beckmann, M.~DeLuca, J.~T. Devlin, and S.~M. Smith, ``Investigations into
  resting-state connectivity using independent component analysis,''
  \emph{Philosophical Transactions of the Royal Society B: Biological
  Sciences}, vol. 360, no. 1457, pp. 1001--1013, 2005.

\bibitem{eavani2012sparse}
H.~Eavani, R.~Filipovych, C.~Davatzikos, T.~D. Satterthwaite, R.~E. Gur, and
  R.~C. Gur, ``Sparse dictionary learning of resting state fmri networks,'' in
  \emph{2012 Second International Workshop on Pattern Recognition in
  NeuroImaging}.\hskip 1em plus 0.5em minus 0.4em\relax IEEE, 2012, pp. 73--76.

\bibitem{bullmore2009complex}
E.~Bullmore and O.~Sporns, ``Complex brain networks: graph theoretical analysis
  of structural and functional systems,'' \emph{Nature reviews neuroscience},
  vol.~10, no.~3, p. 186, 2009.

\bibitem{wang2013analysis}
Y.~Wang and T.-Q. Li, ``Analysis of whole-brain resting-state fmri data using
  hierarchical clustering approach,'' \emph{PloS one}, vol.~8, no.~10, p.
  e76315, 2013.

\bibitem{liu2012correlation}
X.~Liu, X.-H. Zhu, P.~Qiu, and W.~Chen, ``A correlation-matrix-based
  hierarchical clustering method for functional connectivity analysis,''
  \emph{Journal of neuroscience methods}, vol. 211, no.~1, pp. 94--102, 2012.

\bibitem{ashourvan2017multi}
A.~Ashourvan, Q.~K. Telesford, T.~Verstynen, J.~M. Vettel, and D.~S. Bassett,
  ``Multi-scale detection of hierarchical community architecture in structural
  and functional brain networks,'' \emph{arXiv preprint arXiv:1704.05826},
  2017.

\bibitem{betzel2017multi}
R.~F. Betzel and D.~S. Bassett, ``Multi-scale brain networks,''
  \emph{Neuroimage}, vol. 160, pp. 73--83, 2017.

\bibitem{puxeddu2020modular}
M.~G. Puxeddu, J.~Faskowitz, R.~F. Betzel, M.~Petti, L.~Astolfi, and O.~Sporns,
  ``The modular organization of brain cortical connectivity across the human
  lifespan,'' \emph{NeuroImage}, p. 116974, 2020.

\bibitem{betzel2015functional}
R.~F. Betzel, B.~Mi{\v{s}}i{\'c}, Y.~He, J.~Rumschlag, X.-N. Zuo, and
  O.~Sporns, ``Functional brain modules reconfigure at multiple scales across
  the human lifespan,'' \emph{arXiv preprint arXiv:1510.08045}, 2015.

\bibitem{xu2016large}
J.~Xu, M.~N. Potenza, V.~D. Calhoun, R.~Zhang, S.~W. Yip, J.~T. Wall, G.~D.
  Pearlson, P.~D. Worhunsky, K.~A. Garrison, and J.~M. Moran, ``Large-scale
  functional network overlap is a general property of brain functional
  organization: reconciling inconsistent fmri findings from
  general-linear-model-based analyses,'' \emph{Neuroscience \& Biobehavioral
  Reviews}, vol.~71, pp. 83--100, 2016.

\bibitem{lancichinetti2009detecting}
A.~Lancichinetti, S.~Fortunato, and J.~Kertesz, ``Detecting the overlapping and
  hierarchical community structure in complex networks,'' \emph{New Journal of
  Physics}, vol.~11, no.~3, p. 033015, 2009.

\bibitem{shen2009detect}
H.~Shen, X.~Cheng, K.~Cai, and M.-B. Hu, ``Detect overlapping and hierarchical
  community structure in networks,'' \emph{Physica A: Statistical Mechanics and
  its Applications}, vol. 388, no.~8, pp. 1706--1712, 2009.

\bibitem{lancichinetti2011finding}
A.~Lancichinetti, F.~Radicchi, J.~J. Ramasco, and S.~Fortunato, ``Finding
  statistically significant communities in networks,'' \emph{PloS one}, vol.~6,
  no.~4, p. e18961, 2011.

\bibitem{zhang2015mining}
Z.~Zhang and Z.~Wang, ``Mining overlapping and hierarchical communities in
  complex networks,'' \emph{Physica A: Statistical Mechanics and its
  Applications}, vol. 421, pp. 25--33, 2015.

\bibitem{rubinov2011weight}
M.~Rubinov and O.~Sporns, ``Weight-conserving characterization of complex
  functional brain networks,'' \emph{Neuroimage}, vol.~56, no.~4, pp.
  2068--2079, 2011.

\bibitem{zhan2017significance}
L.~Zhan, L.~M. Jenkins, O.~E. Wolfson, J.~J. GadElkarim, K.~Nocito, P.~M.
  Thompson, O.~A. Ajilore, M.~K. Chung, and A.~D. Leow, ``The significance of
  negative correlations in brain connectivity,'' \emph{Journal of Comparative
  Neurology}, vol. 525, no.~15, pp. 3251--3265, 2017.

\bibitem{fox2005human}
M.~D. Fox, A.~Z. Snyder, J.~L. Vincent, M.~Corbetta, D.~C. Van~Essen, and M.~E.
  Raichle, ``The human brain is intrinsically organized into dynamic,
  anticorrelated functional networks,'' \emph{Proceedings of the National
  Academy of Sciences}, vol. 102, no.~27, pp. 9673--9678, 2005.

\bibitem{fitzpatrick2007associations}
A.~L. Fitzpatrick, C.~K. Buchanan, R.~L. Nahin, S.~T. DeKosky, H.~H. Atkinson,
  M.~C. Carlson, and J.~D. Williamson, ``Associations of gait speed and other
  measures of physical function with cognition in a healthy cohort of elderly
  persons,'' \emph{The Journals of Gerontology Series A: Biological Sciences
  and Medical Sciences}, vol.~62, no.~11, pp. 1244--1251, 2007.

\bibitem{goulas2015strength}
A.~Goulas, A.~Schaefer, and D.~S. Margulies, ``The strength of weak connections
  in the macaque cortico-cortical network,'' \emph{Brain Structure and
  Function}, vol. 220, no.~5, pp. 2939--2951, 2015.

\bibitem{santarnecchi2014efficiency}
E.~Santarnecchi, G.~Galli, N.~R. Polizzotto, A.~Rossi, and S.~Rossi,
  ``Efficiency of weak brain connections support general cognitive
  functioning,'' \emph{Human brain mapping}, vol.~35, no.~9, pp. 4566--4582,
  2014.

\bibitem{eavani2015identifying}
H.~Eavani, T.~D. Satterthwaite, R.~Filipovych, R.~E. Gur, R.~C. Gur, and
  C.~Davatzikos, ``Identifying sparse connectivity patterns in the brain using
  resting-state fmri,'' \emph{Neuroimage}, vol. 105, pp. 286--299, 2015.

\bibitem{sahoo2019sparse}
D.~Sahoo, N.~Honnorat, and C.~Davatzikos, ``Sparse low-dimensional causal
  modeling for the analysis of brain function,'' in \emph{Medical Imaging 2019:
  Image Processing}, vol. 10949.\hskip 1em plus 0.5em minus 0.4em\relax
  International Society for Optics and Photonics, 2019, p. 109492R.

\bibitem{sahoo2018gpu}
------, ``Gpu accelerated extraction of sparse granger causality patterns,'' in
  \emph{2018 IEEE 15th International Symposium on Biomedical Imaging (ISBI
  2018)}.\hskip 1em plus 0.5em minus 0.4em\relax IEEE, 2018, pp. 604--607.

\bibitem{yang2013overlapping}
J.~Yang and J.~Leskovec, ``Overlapping community detection at scale: a
  nonnegative matrix factorization approach,'' in \emph{Proceedings of the
  sixth ACM international conference on Web search and data mining}.\hskip 1em
  plus 0.5em minus 0.4em\relax ACM, 2013, pp. 587--596.

\bibitem{song2013hierarchical}
H.~A. Song and S.-Y. Lee, ``Hierarchical representation using nmf,'' in
  \emph{International conference on neural information processing}.\hskip 1em
  plus 0.5em minus 0.4em\relax Springer, 2013, pp. 466--473.

\bibitem{li2018identification}
H.~Li, X.~Zhu, and Y.~Fan, ``Identification of multi-scale hierarchical brain
  functional networks using deep matrix factorization,'' in \emph{International
  Conference on Medical Image Computing and Computer-Assisted
  Intervention}.\hskip 1em plus 0.5em minus 0.4em\relax Springer, 2018, pp.
  223--231.

\bibitem{trigeorgis2017deep}
G.~Trigeorgis, K.~Bousmalis, S.~Zafeiriou, and B.~W. Schuller, ``A deep matrix
  factorization method for learning attribute representations,'' \emph{IEEE
  transactions on pattern analysis and machine intelligence}, vol.~39, no.~3,
  pp. 417--429, 2017.

\bibitem{madsen2017quantifying}
K.~H. Madsen, N.~W. Churchill, and M.~M{\o}rup, ``Quantifying functional
  connectivity in multi-subject fmri data using component models,'' \emph{Human
  brain mapping}, vol.~38, no.~2, pp. 882--899, 2017.

\bibitem{van2013wu}
D.~C. Van~Essen, S.~M. Smith, D.~M. Barch, T.~E. Behrens, E.~Yacoub,
  K.~Ugurbil, W.-M.~H. Consortium \emph{et~al.}, ``The wu-minn human connectome
  project: an overview,'' \emph{Neuroimage}, vol.~80, pp. 62--79, 2013.

\bibitem{satterthwaite2014neuroimaging}
T.~D. Satterthwaite, M.~A. Elliott, K.~Ruparel, J.~Loughead, K.~Prabhakaran,
  M.~E. Calkins, R.~Hopson, C.~Jackson, J.~Keefe, M.~Riley \emph{et~al.},
  ``Neuroimaging of the philadelphia neurodevelopmental cohort,''
  \emph{Neuroimage}, vol.~86, pp. 544--553, 2014.

\bibitem{achard2007efficiency}
S.~Achard and E.~Bullmore, ``Efficiency and cost of economical brain functional
  networks,'' \emph{PLoS computational biology}, vol.~3, no.~2, p. e17, 2007.

\bibitem{podosinnikova2013robust}
A.~Podosinnikova, M.~Hein, and R.~Gemulla, ``Robust principal component
  analysis as a nonlinear eigenproblem,'' Ph.D. dissertation, Saarland
  University, 2013.

\bibitem{reddi2019convergence}
S.~J. Reddi, S.~Kale, and S.~Kumar, ``On the convergence of adam and beyond,''
  \emph{arXiv preprint arXiv:1904.09237}, 2019.

\bibitem{kingma2014adam}
D.~P. Kingma and J.~Ba, ``Adam: A method for stochastic optimization,''
  \emph{arXiv preprint arXiv:1412.6980}, 2014.

\bibitem{dozat2016incorporating}
T.~Dozat, ``Incorporating nesterov momentum into adam,'' 2016.

\bibitem{van2012human}
D.~C. Van~Essen, K.~Ugurbil, E.~Auerbach, D.~Barch, T.~Behrens, R.~Bucholz,
  A.~Chang, L.~Chen, M.~Corbetta, S.~W. Curtiss \emph{et~al.}, ``The human
  connectome project: a data acquisition perspective,'' \emph{Neuroimage},
  vol.~62, no.~4, pp. 2222--2231, 2012.

\bibitem{glasser2013minimal}
M.~F. Glasser, S.~N. Sotiropoulos, J.~A. Wilson, T.~S. Coalson, B.~Fischl,
  J.~L. Andersson, J.~Xu, S.~Jbabdi, M.~Webster, J.~R. Polimeni \emph{et~al.},
  ``The minimal preprocessing pipelines for the human connectome project,''
  \emph{Neuroimage}, vol.~80, pp. 105--124, 2013.

\bibitem{glasser2016multi}
M.~F. Glasser, T.~S. Coalson, E.~C. Robinson, C.~D. Hacker, J.~Harwell,
  E.~Yacoub, K.~Ugurbil, J.~Andersson, C.~F. Beckmann, M.~Jenkinson
  \emph{et~al.}, ``A multi-modal parcellation of human cerebral cortex,''
  \emph{Nature}, vol. 536, no. 7615, pp. 171--178, 2016.

\bibitem{doshi2016muse}
J.~Doshi, G.~Erus, Y.~Ou, S.~M. Resnick, R.~C. Gur, R.~E. Gur, T.~D.
  Satterthwaite, S.~Furth, C.~Davatzikos, A.~N. Initiative \emph{et~al.},
  ``Muse: Multi-atlas region segmentation utilizing ensembles of registration
  algorithms and parameters, and locally optimal atlas selection,''
  \emph{Neuroimage}, vol. 127, pp. 186--195, 2016.

\bibitem{ciric2017benchmarking}
R.~Ciric, D.~H. Wolf, J.~D. Power, D.~R. Roalf, G.~L. Baum, K.~Ruparel, R.~T.
  Shinohara, M.~A. Elliott, S.~B. Eickhoff, C.~Davatzikos \emph{et~al.},
  ``Benchmarking of participant-level confound regression strategies for the
  control of motion artifact in studies of functional connectivity,''
  \emph{Neuroimage}, vol. 154, pp. 174--187, 2017.

\bibitem{wagstaff2001constrained}
K.~Wagstaff, C.~Cardie, S.~Rogers, S.~Schr{\"o}dl \emph{et~al.}, ``Constrained
  k-means clustering with background knowledge,'' in \emph{Icml}, vol.~1, 2001,
  pp. 577--584.

\bibitem{raichle2015brain}
M.~E. Raichle, ``The brain's default mode network,'' \emph{Annual review of
  neuroscience}, vol.~38, pp. 433--447, 2015.

\bibitem{yeo2014estimates}
B.~T. Yeo, F.~M. Krienen, M.~W. Chee, and R.~L. Buckner, ``Estimates of
  segregation and overlap of functional connectivity networks in the human
  cerebral cortex,'' \emph{Neuroimage}, vol.~88, pp. 212--227, 2014.

\bibitem{buckner2013opportunities}
R.~L. Buckner, F.~M. Krienen, and B.~T. Yeo, ``Opportunities and limitations of
  intrinsic functional connectivity mri,'' \emph{Nature neuroscience}, vol.~16,
  no.~7, pp. 832--837, 2013.

\bibitem{karahanouglu2015transient}
F.~I. Karahano{\u{g}}lu and D.~Van De~Ville, ``Transient brain activity
  disentangles fmri resting-state dynamics in terms of spatially and temporally
  overlapping networks,'' \emph{Nature communications}, vol.~6, no.~1, pp.
  1--10, 2015.

\bibitem{drevets1998reciprocal}
W.~C. Drevets and M.~E. Raichle, ``Reciprocal suppression of regional cerebral
  blood flow during emotional versus higher cognitive processes: Implications
  for interactions between emotion and cognition,'' \emph{Cognition and
  emotion}, vol.~12, no.~3, pp. 353--385, 1998.

\bibitem{vergani2014white}
F.~Vergani, L.~Lacerda, J.~Martino, J.~Attems, C.~Morris, P.~Mitchell, M.~T.
  de~Schotten, and F.~Dell'Acqua, ``White matter connections of the
  supplementary motor area in humans,'' \emph{Journal of Neurology,
  Neurosurgery \& Psychiatry}, vol.~85, no.~12, pp. 1377--1385, 2014.

\bibitem{geranmayeh2014overlapping}
F.~Geranmayeh, R.~J. Wise, A.~Mehta, and R.~Leech, ``Overlapping networks
  engaged during spoken language production and its cognitive control,''
  \emph{Journal of Neuroscience}, vol.~34, no.~26, pp. 8728--8740, 2014.

\bibitem{beldzik2013contributive}
E.~Beldzik, A.~Domagalik, S.~Daselaar, M.~Fafrowicz, W.~Froncisz, H.~Oginska,
  and T.~Marek, ``Contributive sources analysis: a measure of neural networks'
  contribution to brain activations,'' \emph{Neuroimage}, vol.~76, pp.
  304--312, 2013.

\bibitem{katsuki2012unique}
F.~Katsuki and C.~Constantinidis, ``Unique and shared roles of the posterior
  parietal and dorsolateral prefrontal cortex in cognitive functions,''
  \emph{Frontiers in integrative neuroscience}, vol.~6, p.~17, 2012.

\bibitem{crossley2014hubs}
N.~A. Crossley, A.~Mechelli, J.~Scott, F.~Carletti, P.~T. Fox, P.~McGuire, and
  E.~T. Bullmore, ``The hubs of the human connectome are generally implicated
  in the anatomy of brain disorders,'' \emph{Brain}, vol. 137, no.~8, pp.
  2382--2395, 2014.

\bibitem{cai2017estimation}
B.~Cai, P.~Zille, J.~M. Stephen, T.~W. Wilson, V.~D. Calhoun, and Y.~P. Wang,
  ``Estimation of dynamic sparse connectivity patterns from resting state
  fmri,'' \emph{IEEE transactions on medical imaging}, vol.~37, no.~5, pp.
  1224--1234, 2017.

\bibitem{yu2018recovery}
M.~Yu, Z.~Wang, V.~Gupta, and M.~Kolar, ``Recovery of simultaneous low rank and
  two-way sparse coefficient matrices, a nonconvex approach,'' \emph{arXiv
  preprint arXiv:1802.06967}, 2018.

\end{thebibliography}


\begin{thebibliography}{10}

\bibitem{ashourvan2017multi}
Arian Ashourvan, Qawi~K Telesford, Timothy Verstynen, Jean~M Vettel, and
  Danielle~S Bassett.
\newblock Multi-scale detection of hierarchical community architecture in
  structural and functional brain networks.
\newblock {\em arXiv preprint arXiv:1704.05826}, 2017.

\bibitem{beckmann2005investigations}
Christian~F Beckmann, Marilena DeLuca, Joseph~T Devlin, and Stephen~M Smith.
\newblock Investigations into resting-state connectivity using independent
  component analysis.
\newblock {\em Philosophical Transactions of the Royal Society B: Biological
  Sciences}, 360(1457):1001--1013, 2005.

\bibitem{bullmore2009complex}
Ed~Bullmore and Olaf Sporns.
\newblock Complex brain networks: graph theoretical analysis of structural and
  functional systems.
\newblock {\em Nature reviews neuroscience}, 10(3):186, 2009.

\bibitem{dozat2016incorporating}
Timothy Dozat.
\newblock Incorporating nesterov momentum into adam.
\newblock 2016.

\bibitem{eavani2012sparse}
Harini Eavani, Roman Filipovych, Christos Davatzikos, Theodore~D Satterthwaite,
  Raquel~E Gur, and Ruben~C Gur.
\newblock Sparse dictionary learning of resting state fmri networks.
\newblock In {\em 2012 Second International Workshop on Pattern Recognition in
  NeuroImaging}, pages 73--76. IEEE, 2012.

\bibitem{eavani2015identifying}
Harini Eavani, Theodore~D Satterthwaite, Roman Filipovych, Raquel~E Gur,
  Ruben~C Gur, and Christos Davatzikos.
\newblock Identifying sparse connectivity patterns in the brain using
  resting-state fmri.
\newblock {\em Neuroimage}, 105:286--299, 2015.

\bibitem{ferrarini2009hierarchical}
Luca Ferrarini, Ilya~M Veer, Evelinda Baerends, Marie-Jos{\'e} van Tol, Remco~J
  Renken, Nic~JA van~der Wee, Dirk~J Veltman, Andre Aleman, Frans~G Zitman,
  Brenda~WJH Penninx, et~al.
\newblock Hierarchical functional modularity in the resting-state human brain.
\newblock {\em Human brain mapping}, 30(7):2220--2231, 2009.

\bibitem{golub1987generalization}
Gene~H Golub, Alan Hoffman, and Gilbert~W Stewart.
\newblock A generalization of the eckart-young-mirsky matrix approximation
  theorem.
\newblock {\em Linear Algebra and its applications}, 88:317--327, 1987.

\bibitem{kingma2014adam}
Diederik~P Kingma and Jimmy Ba.
\newblock Adam: A method for stochastic optimization.
\newblock {\em arXiv preprint arXiv:1412.6980}, 2014.

\bibitem{li2018identification}
Hongming Li, Xiaofeng Zhu, and Yong Fan.
\newblock Identification of multi-scale hierarchical brain functional networks
  using deep matrix factorization.
\newblock In {\em International Conference on Medical Image Computing and
  Computer-Assisted Intervention}, pages 223--231. Springer, 2018.

\bibitem{liu2012correlation}
Xiao Liu, Xiao-Hong Zhu, Peihua Qiu, and Wei Chen.
\newblock A correlation-matrix-based hierarchical clustering method for
  functional connectivity analysis.
\newblock {\em Journal of neuroscience methods}, 211(1):94--102, 2012.

\bibitem{meunier2009hierarchical}
David Meunier, Renaud Lambiotte, Alex Fornito, Karen Ersche, and Edward~T
  Bullmore.
\newblock Hierarchical modularity in human brain functional networks.
\newblock {\em Frontiers in neuroinformatics}, 3:37, 2009.

\bibitem{podosinnikova2013robust}
Anastasia Podosinnikova, Matthias Hein, and Rainer Gemulla.
\newblock {\em Robust Principal Component Analysis as a Nonlinear
  Eigenproblem}.
\newblock PhD thesis, Saarland University, 2013.

\bibitem{reddi2018convergence}
Sashank~J Reddi, Satyen Kale, and Sanjiv Kumar.
\newblock On the convergence of adam and beyond.
\newblock 2018.

\bibitem{rosvall2008maps}
Martin Rosvall and Carl~T Bergstrom.
\newblock Maps of random walks on complex networks reveal community structure.
\newblock {\em Proceedings of the National Academy of Sciences},
  105(4):1118--1123, 2008.

\bibitem{satterthwaite2014neuroimaging}
Theodore~D Satterthwaite, Mark~A Elliott, Kosha Ruparel, James Loughead,
  Karthik Prabhakaran, Monica~E Calkins, Ryan Hopson, Chad Jackson, Jack Keefe,
  Marisa Riley, et~al.
\newblock Neuroimaging of the philadelphia neurodevelopmental cohort.
\newblock {\em Neuroimage}, 86:544--553, 2014.

\bibitem{trigeorgis2017deep}
George Trigeorgis, Konstantinos Bousmalis, Stefanos Zafeiriou, and Bj{\"o}rn~W
  Schuller.
\newblock A deep matrix factorization method for learning attribute
  representations.
\newblock {\em IEEE transactions on pattern analysis and machine intelligence},
  39(3):417--429, 2017.

\bibitem{wang2013analysis}
Yanlu Wang and Tie-Qiang Li.
\newblock Analysis of whole-brain resting-state fmri data using hierarchical
  clustering approach.
\newblock {\em PloS one}, 8(10):e76315, 2013.

\end{thebibliography}




\end{document}